\documentclass[12pt]{article}

\usepackage[top=1.2in, bottom=1.2in, left=1in, right=1in]{geometry}

\usepackage[sort&compress,authoryear]{natbib}
\usepackage{BOONDOX-cal}
\usepackage{epsfig, color}
\usepackage{graphicx}
\usepackage{amsmath,amsthm,amsfonts,amssymb,bm}
\usepackage[figuresright]{rotating}

\usepackage{makeidx}
\usepackage{listings}
\usepackage[colorlinks,
linkcolor=black,
anchorcolor=black,
citecolor=blue
]{hyperref}
\usepackage{enumerate}
\usepackage{enumitem}

\usepackage{rotating}
\usepackage{graphicx}
\usepackage{tikz}
\usepackage{multirow}
\usepackage{subcaption}
\usepackage{cases}
\newtheorem{remark}{Remark}[section]

\newtheorem{theorem}{\hspace{6mm}Theorem}[section]
\newtheorem{assumption}{\hspace{6mm}Assumption}[section]

\newtheorem{proposition}{\hspace{6mm}Proposition}[section]

\newtheorem{corollary}{\hspace{6mm}Corollary}[section]
\newtheorem{definition}{\hspace{6mm}Definition}[section]

\allowdisplaybreaks[4]

\title{Time-consistent catastrophe risk management under the path-dependent effects}
\author{
		Liyuan Cui\thanks{\rm  Corresponding author. Department of Statistics and Actuarial Science, The University of Hong Kong, Pokfulam, Hong Kong, China. Email: cuiliyuan@connect.hku.hk}
\and
Wenyuan Li\thanks{\rm Department of Statistics and Actuarial Science, The University of Hong Kong, Pokfulam, Hong Kong, China. Email: wylsaas@hku.hk} 
	}
\date{\today}

\begin{document}

\maketitle

\begin{abstract}
This paper investigates optimal investment and insurance strategies under a mean-
variance criterion with path-dependent effects. We use a rough volatility model with a power kernel and a Hawkes process with a modified Omori kernel (a power kernel) to capture the market's path dependence. By extending the functional Ito calculus to the mixed fractional Brownian-Hawkes process, we derive the corresponding path-dependent extended Hamilton-Jacobi-Bellman equation and solve its explicit solution. For numerical analysis, we first calibrate the Hawkes process with the Wenchuan earthquake data, the most devastating earthquake in China. We find that the power kernel outperforms the exponential one in fitting the earthquake intensity. Our numerical results reveal that the path-dependent effect strongly depends on the horizon length. For the investment strategy, the individual is more risk-seeking when considering the rougher volatility in the short horizon, but more risk-averse at the beginning of the long horizon. For the insurance strategy, a faster decay in intensity increases individuals’ demand for catastrophe insurance in the short horizon, but decreases initially in the long horizon. However, we find that the horizon effect may disappear when the shift parameter in the modified Omori kernel approaches zero or exceeds one. Our findings indicate that ignoring path-dependent effects would lead to significant underinsurance and highlight its importance in catastrophe risk management. 
\end{abstract}

\clearpage
\section{Introduction}
The optimal investment-insurance problem under the mean-variance criterion has been a central focus of research in actuarial science. Recent studies have explored this problem through various frameworks, such as insurers’ aversion to model uncertainty (\cite{1Zeng2016robust}, \cite{2Zhao2019RobustEE}, and \cite{20Zeng2013TimeconsistentIA}), claim contagion models expanding on the Cox and Hawkes processes (\cite{4Dassios} and \cite{3Cao2020OptimalRS}), wealth processes with bankruptcy constraints (\cite{5Bi}), catastrophic and secondary claims models (\cite{6Zhang}), and proportional reinsurance with group correlation (\cite{7Pressacco2011MeanvarianceES}). Under the mean-variance criterion, time inconsistency is a common issue due to the objective's inherent quadratic structure (see \cite{N_time_inconsistent} and \cite{Bjork_2010}). To address this fundamental problem, the extended Hamilton-Jacobi-Bellman (HJB) equation is introduced by seeking Nash equilibrium strategies within a non-cooperative game-theoretic framework. For related works, we refer to \cite{20Zeng2013TimeconsistentIA}, \\ \cite{O_time_inconsistent}, \cite{N_time_inconsistent}, and \cite{P_time_inconsistent}. In the current literature, it is common to assume that market volatility and jump processes follow a Markov or semimartingale structure. However, empirical evidence indicates that stock volatility and catastrophe events, such as earthquakes and tsunamis, exhibit strong path dependence that is not adequately captured by the classical Markovian framework. 

Path dependence in stock volatility means that volatility's future dynamics are shaped not only by the current state but also by its historical state. \cite{8Gatheral2014VolatilityIR} first show that major financial indices exhibit a strong path-dependent feature and propose a rough volatility model. Subsequently, numerous papers have considered optimal decision-making under the rough volatility model, including \cite{9Bayer2015PricingUR}, \cite{10article}, \cite{11article}, and \cite{12Han2021RobustCI}. However, they study only optimal investment decisions, leaving research gaps regarding optimal insurance demand. In actuarial science, catastrophe shocks, such as earthquakes and tsunamis, exhibit a significant path dependence: Large shocks are always followed by smaller ones. The Hawkes process, as a self-exciting point process, is well suited to capturing such clustering behavior. Unlike the memoryless Poisson process, it allows the conditional intensity to depend on past jumps, so that more previous jumps lead to a higher current intensity, reflecting self-exciting or contagious effects. In a typical Hawkes process, past effects decay over time through a kernel function. Empirical studies show that the power kernel outperforms the exponential kernel in describing aftershock effects. Specifically, in the seismological literature, \cite{Omori1984} first observes an inverse relationship between aftershock intensity and time, known as Omori’s Law. Later, \cite{utsu1961statistical} proposes a modified Omori’s Law where aftershock intensity is a power function of time and better fits the earthquake data. After the seminal work \cite{15Hawkes1971SpectraOS}, \cite{adamopoulos1976cluster} first introduces the Hawkes process with an exponential kernel to fit aftershock data. Later, \cite{I_Hawkes} and \cite{1.hawkes_Ogata} incorporate the Hawkes process with the modified Omori’s Law, yielding the Epidemic Type Aftershock-Sequences model. They use log-likelihood estimation to compare different models and find that the Hawkes process with a power kernel outperforms the one with an exponential kernel. Subsequent studies support this conclusion and continue to improve the ETAS model (see \cite{ross2022semiparametric}, \cite{davis2024fractional}, and \cite{mizrahi2024developing}). In contrast, existing actuarial research typically employs an exponential kernel because it is memoryless and can reduce a non-Markovian problem to a Markovian one (see \cite{dassios2012ruin}, \cite{3Cao2020OptimalRS}, \cite{brachetta2024optimal}, \cite{6Zhang}, and \cite{feng2025robust}). This discrepancy motivates us to study the demand for catastrophe insurance under the Hawkes process with a power kernel rather than an exponential kernel. It is also challenging to achieve this objective, as the Hawkes process under the power kernel is a fully non-Markovian model and requires new techniques for its solution. 

This paper investigates the role of path-dependent effects in the optimal investment-insurance problem. Specifically, we consider an individual who simultaneously manages the catastrophe and financial risks. In the financial market, we model the stock volatility using the Volterra Heston framework from \cite{bacry2015hawkes}, which captures the rough volatility property widely observed in empirical data. In the insurance market, we model catastrophe risk using a Hawkes process with a power kernel, consistent with the modified Omori’s Law, to capture the clustering and decay of aftershock activity. The individual’s objective is to maximize expected terminal utility under a time-consistent mean–variance criterion. To solve the problem, we first derive the corresponding extended HJB equations. Since both the Volterra process and the Hawkes process are inherently non-Markovian and non-semimartingale, we employ functional Itô calculus from \cite{21article} to address these challenges, leading to explicit expressions for the equilibrium investment–insurance strategy and the associated value function.

We calibrate our model using earthquake data from Sichuan Province in China from 2008 to 2023, focusing on events with a magnitude above 5.0 (moderate earthquakes that can cause economic losses), and numerically demonstrate the impact of path dependence on the individual’s investment and insurance decisions. In the investment problem, we find that rough volatility effects on investment are highly related to the investment horizon. Specifically, over short-term horizons, the individual becomes more risk-seeking under the rougher volatility model. In contrast, over long-term horizons, individuals initially reduce stock allocations when considering the rough volatility, reflecting a reversal driven by the horizon documented in Corollary 4.1 of \cite{HanWong}. We also identify a significant horizon effect regarding insurance demand. In the short term, a faster decay in intensity leads to cheaper premiums under a fixed severity, thereby increasing individuals' demand for catastrophe insurance. Over the long term, demand decreases as the intensity decays more rapidly, though this trend eventually reverses toward the short-term pattern as maturity approaches. In contrast to the existing literature, we find that the horizon effect may disappear for catastrophe insurance when the shift parameter in the modified Omori kernel approaches zero or exceeds one. Besides the horizon effect, the individual generally purchases less insurance under the standard Poisson intensity than under the Hawkes intensity, suggesting that ignoring the path-dependent effect can lead to underinsurance.

Our paper contributes to the literature on optimal investment–insurance by incorporating path-dependent effects into both the financial and insurance markets. Our contributions can be summarized in three aspects: (1) We extend the exponential kernel Hawkes framework to a power kernel, thereby integrating actuarial catastrophe modeling with empirical findings from seismology. This allows us to capture the long-memory and clustering effects of catastrophic events more realistically. (2) We extend the Functional Itô calculus framework of fractional Brownian motion to the mixed fractional Brownian-Hawkes process. This allows us to derive and solve extended HJB equations under non-Markovian and non-semimartingale dynamics. (3) Our results show that the path-dependent effect is closely tied to horizon length. In investment, individuals are more risk-seeking in the short term with rougher volatility, but more risk-averse at the start of the long term. In insurance, the horizon effect may disappear when the shift parameter in the modified Omori kernel approaches zero or exceeds one.

Specifically, underinsurance is a phenomenon where an individual doesn't purchase enough insurance to cover their risks. \cite{lloyds2018worldatrisk} shows a significant insurance gap in the global market, amounting to around US $\$2.5$ billion in developed countries and US $\$160$ billion in emerging nations. The existing literature explains the underinsurance phenomenon through market structures and individuals' preferences, such as intertemporal considerations like habit formation (\cite{1.underinsurance}), behavioral biases like narrow framing and skewness preferences (\cite{2.underinsurance}), and monopolistic pricing in Stackelberg equilibria (\cite{3.underinsurance}). Our research indicates that path-dependent effects of natural processes, such as aftershock clustering in earthquakes and tsunamis, are also a potential driver of underinsurance. We suggest that insurance companies educate the public about the path-dependent effects of catastrophe risk to increase voluntary purchases of catastrophe insurance.

The remainder of this paper is organized as follows. Section \ref{sec:model} introduces the formulation of the problem. Section \ref{section3:functional formula} extends the functional Itô formula to the mixed fractional Brownian-Hawkes process. Section \ref{sec:solution} derives explicit solutions for the equilibrium investment-insurance strategy and the corresponding value function. Section \ref{sec:numerical} presents the model calibration results and the numerical simulation of the problem. Section \ref{sec:conclusion} concludes the paper. Technical proofs are provided in the appendix.

\section{Economic settings}\label{sec:model}
Our problem is defined under a given complete probability space $(\Omega, \mathcal{F}, \mathbb{P})$, with a filtration
$\mathbb{F} = \left \{ \mathcal{F}_t  \right \}_{0\le t\le T}$ satisfying the usual conditions. One can separate the filtration $\mathcal{F}_t=\mathcal{F}^W_t \vee \mathcal{H}_t$, where $\mathcal{F}^W_t$ is a filtration generated by a two-dimensional Brownian motion
 $W =(W_1, W_2)$ and $\mathcal{H}_t$ is filtration generated by a compounded Hawkes process $Y_t$. 
\subsection{Financial market}
In the financial market, we assume the stock volatility follows the Volterra Heston model (see \cite{Abi2019lifting})
 \begin{equation}\label{volatility_1}
     v_t = v_0 + \kappa \int_{0}^{t} K(t-r)(\phi - v_r) dr + \int_{0}^{t} K(t-r) \sigma \sqrt{v_r} d\textbf{B}_r,
 \end{equation}
 where $K(\cdot) \in L^2_{loc}(\mathbb{R}_{\ge 0}, \mathbb{R})$ is the kernel function, $d\textbf{B}_r = \rho dW_{1r}+ \sqrt{1-\rho^2}dW_{2r}$ with $\rho \in \left [ -1,1 \right ]$, and $v_0, \kappa, \phi, \sigma$ are positive constants. In this paper, we assume a power kernel function $K(t)=t^{\delta-1}/\Gamma(\delta),\,\, \delta \in (\frac{1}{2},1]$ following \cite{abi2019affine}, \cite{wang2021volterra}, and \cite{wang2021time}. Here, $\delta$ is a parameter measuring the long-range dependence of the stock volatility, which is related to the Hurst parameter $H_{Hurst} = \delta - 1/2$ in the rough theory. In particular, a smaller $\delta$ corresponds to a smaller Hurst parameter, leading to a rougher volatility path. When $\delta = 1$, the Volterra Heston model is degenerated to a classical Heston model. Inspired by \cite{HanWong} and \cite{12Han2021RobustCI}, we assume the stock price $S_t$ under the physical measure $\mathbb{P}$ is given by
 \begin{equation*}
     dS_t = S_t(\Upsilon + \theta v_t)dt + S_t\sqrt{v_t}dW_{1t}, \,\,  S_0 > 0,
 \end{equation*}
 with a deterministic risk-free rate $\Upsilon > 0$ and constant $\theta \ne 0$. The market price of risk, or risk premium, is then given by $\theta \sqrt{v_t}$. Compared to the classical Heston model, the Volterra Heston model can better capture large price movements near maturity and better match the implied volatility surface (see and \cite{9Bayer2015PricingUR}, \cite{Abi2019lifting}, and \cite{el2019characteristic}).

 \subsection{Insurance market}
 \subsubsection{Representation of compounded Hawkes process}

 In the insurance market, we assume the individual's risk is modeled by a compounded Hawkes process
\begin{equation}\label{compounded hawkes process}
    Y_t = \sum \limits_{i=1}^{\pi_t} \widetilde{Y}_i,
\end{equation}
where $\left \{ \pi_t\right \}_{0\le t\le T}$ is a Hawkes process representing the number of losses up to time t and the $\widetilde{Y}_i$,  $i\in \mathbb{Z}^+$ is a sequence of i.i.d. random variables taking values in $\mathbb{R}^+$ with common distribution $F_{\widetilde{Y}}$. 

To characterize a Hawkes process, one can specify the distribution of the next arrival time by the conditional density. Following \cite{2.hawkes_Patrik}, we let $\mathcal{H}_t$ denote the filtration of arrivals up to time $t$. Consequently, the conditional intensity function is the expected rate of future arrivals conditioned on $\mathcal{H}_t$.
\begin{definition}
    \textbf{(Conditional intensity function)} Consider a counting process $\pi_{t}$ with associated filtration $\mathcal{H}_t$. If a (non-negative) function $\lambda_t$ exists such that
 \begin{equation*}
     \lambda_t=\lim_{h\to 0}\frac{\mathbb{E}\left[\pi_{t+h}-\pi_t|\mathcal{H}_t\right]}{h},
 \end{equation*}
  which relies only on information from $\mathcal{H}_t$ in the past, then it is called the conditional intensity function of $\pi_t$.
\end{definition}

Among the counting processes, the Hawkes process is uniquely characterized by a stochastic conditional intensity that is linear in past jumps. Specifically, the conditional intensity of $\pi_t$ in \eqref{compounded hawkes process} is given by
\begin{equation}\label{intensity_1}
\lambda_t=\lambda_0+\int_{0}^{t}\varphi (t-r)(a_0+a_1\lambda_r)dr+\int_{0}^{t}\varphi (t-r)d\pi_r,
\end{equation}
where $a_0\ge 0$, $a_1\in(-\infty,+\infty)$, $\lambda_0$ is the baseline intensity, and $\varphi$ is a positive kernel function belonging to the  $L_{loc}^1-$integrable functions. We take $\varphi(t)=\varrho_1/\left(\varrho_2+t\right)^p$ with some positive scalars $\varrho_1, \varrho_2, p$ to capture the path dependence effect of jump process. In particular, a larger $p$ corresponds to an intensity path that decays more quickly. This power-law form, called modified Omori's law (see \cite{Utsu1995}, \cite{I_Hawkes}, and \cite{1.hawkes_Ogata}), is used to describe the aftershock clustering among earthquakes. For comparison, we also consider the exponential kernel function as $\varphi(t)=b_1e^{-b_2 t}$, where $b_1, b_2$ are positive constants. Due to the memoryless/semigroup property of the exponential kernel, the control problem degenerates to a Markovian problem.

\subsubsection{Compensator of compounded Hawkes process}
According to the definition of the conditional intensity function, we know
\begin{equation*}
\mathbb{E}\left[\pi_t\right]=\int_0^t\mathbb{E}\left[\lambda_s\right]ds.
\end{equation*}
Let $B_0$ denote the family of Borel sets $U\in \mathbb{R}$ whose closure does not contain $0$. Inspired by \cite{delattre2016hawkes} and \cite{jumpbook}, for $U\in B_0$, the poisson random measure $N(U, \lambda, t)$ of $Y_t$ in \eqref{compounded hawkes process} satisfies
\begin{equation*}
        N(U, \lambda, t)=\sum_{0\le s \le t}\textbf{1}_{U}(\triangle Y_s)=\int_0^t \int_0^{\infty}\int_U\textbf{1}_{\left \{ \lambda\le\lambda_s \right \} }(\lambda)N(dy, d\lambda, ds),
\end{equation*}
where the inner integral integrates over $y$, the middle integral integrates over $\lambda$, and outer integral integrates over time $s$. Moreover, $\textbf{1}_{U}(A)$ is the indicator function. It equals $1$ when $A \in U$ and equals $0$ otherwise. Subsequently, the compensator $\nu$ of $Y_t$ in \eqref{compounded hawkes process} can be calculated as
\begin{eqnarray*}
        \nu(U|\lambda_t, t)&=&\lim_{\triangle t\to 0}\mathbb{E}_t\left[\frac{N(U,\lambda_{t+\Delta t},t+\triangle t)-N(U, \lambda_t,t)}{\Delta t}\right]\\
        &=&\lim_{\triangle t\to 0}\mathbb{E}_t\left[\frac{1}{\Delta t}\sum_{t\le s \le t+\triangle t}\textbf{1}_{U}(\triangle Y_s)\right]\\
        &=&\lim_{\triangle t\to 0}\mathbb{E}_t\left[ \textbf{1}_{U} (\widetilde{Y}) \cdot \frac{\pi_{t+\triangle t}-\pi_t}{\Delta t}\right]\\
        &=&\mathbb{E}_t\left[ \textbf{1}_{U} (\widetilde{Y}) \right]\cdot \lim_{\triangle t\to 0}\mathbb{E}_t\left[\frac{\pi_{t+\triangle t}-\pi_t}{\Delta t}\right]\\
        &=&\mu_{\widetilde{Y}}(U)\cdot\lambda_t = \mu_{\widetilde{Y}}(U) \int_0^{\infty}\textbf{1}_{\{\lambda \leq \lambda_t\}}(\lambda) d\lambda,
\end{eqnarray*}
where the third and forth equality hold true because intensity $\lambda_t$ and severity $\widetilde{Y}$ are independent. $\mu_{\widetilde{Y}}(U)$ is the distribution measure of $\widetilde{Y}$. To obtain the differential form of the compensator, we need to further take the differential of $\nu$ with $\lambda$, i.e., $\nu(U|d\lambda, t)=\mu_{\widetilde{Y}}(U)d\lambda$. Finally, $\widetilde{N}(dy, d\lambda, dt )=N(dy, d\lambda , dt)-\nu(dy|d\lambda, t) dt = N(dy, d\lambda , dt)-\mu_{\widetilde{Y}}(dy)d\lambda dt$ is the compensated Poisson random measure of $Y_t$. Furthermore, based on \cite{delattre2016hawkes} and \cite{jumpbook}, for $\forall U\in B_0, \widetilde{N}(U, \lambda, t )$ is a local martingale. 

To guarantee the Hawkes' compensated Poisson measure is a martingale, we need the subcriticality condition on the kernel $\varphi(t)$ in \eqref{intensity_1}. Specifically, we first rewrite the compensated Hawkes process in an integral form
\begin{eqnarray}
   && \int_0^t\int_0^{\infty} \int_0^{\infty} \textbf{1}_{\{\lambda \leq \lambda_s\}}(\lambda) \widetilde{N}(dy, d\lambda , ds)\notag\\
   &=&\int_0^t \int_0^{\infty} \int_0^{\infty} \textbf{1}_{\{\lambda \leq \lambda_s\}}(\lambda) N(dy, d\lambda , ds)-\int_0^t\int_0^{\infty}  \int_0^{\infty} \textbf{1}_{\{\lambda \leq \lambda_s\}}(\lambda) \mu_{\widetilde{Y}}(dy)d\lambda ds\notag\\
   &=&\int_0^t \int_0^{\infty} \int_0^{\infty} \textbf{1}_{\{\lambda \leq \lambda_s\}}(\lambda) N(dy, d\lambda , ds)- E[\widetilde{Y}]\int_0^t \lambda_s ds,\label{compensated hawkes}
\end{eqnarray}
where the inner integral integrates over $y$, the middle integral integrates over $\lambda$, and the outer integral integrates over time $s$. To guarantee that the compensated Hawkes process above is a true martingale, we need the following proposition 
\begin{proposition}\label{proposition_2.1}
    For the compounded Hawkes process in \eqref{compounded hawkes process} with intensity process in \eqref{intensity_1}, assume that

    (i) the jump severity has finite first moment,
\begin{equation*}
    \mathbb{E}[\widetilde Y]<\infty;
\end{equation*}

(ii) the Hawkes feedback is subcritical, namely
\begin{equation*}
    (a_1+1)_+\,\|\varphi\|_{L^1(0,\infty)}<1,
\end{equation*}
where $(a_1+1)_+:=\max\{a_1+1,0\}$ and $\|\cdot\|_{L^1(0,\infty)}$ is the $L_1$ norm on $(0,\infty)$. Consequently, the compensated compound Hawkes process is a true martingale on every finite horizon $[0,T]$. In particular, for a modified Omori kernel, $\varphi(t)=\varrho_1/(\varrho_2+t)^p$, condition (ii) is implied by
\begin{equation}\label{subcritical condition}
    p>1, ~\text{and}~\frac{\varrho_1(a_1+1)_+}{(p-1)\varrho_2^{p-1}}<1.
\end{equation}
For an exponential kernel, $\varphi(t)=b_1e^{-b_2 t}$, condition (ii) is implied by
\begin{equation}\label{subcritical condition_exponential}
    \frac{b_1(a_1+1)_+}{b_2}<1.
\end{equation}
\begin{proof}
        The detailed proof is in Appendix \ref{proof of proposition_2.1}. For the literature on the subcriticality condition, we refer to \cite{hawkes1974cluster} and \cite{bremaud1996stability}. 
    \end{proof}
\end{proposition}

\subsection{Wealth process and objective function}
We let $\alpha_t$ be a self-financing investment strategy such that $\alpha_t/\sqrt{v_t}$ is the proportion of wealth invested in the stock. The individual is also allowed to purchase per-claim insurance to cover their catastrophic loss $Y$. In particular, denote the individual’s indemnity by $I_t(Y)$, and then the insurance company charges $\lambda_t(1+\widetilde{\theta})\mathbb{E}_t\left[I_t(Y)\right]$ as the premium and pays $I_t(Y)$ if the individual suffers a loss of $Y$ at time $t$. Consequently, the wealth process $X_t$ satisfies 
\begin{equation}\label{wealth process}
    \begin{aligned}
        dX_t=&  \left( \Upsilon + \theta \sqrt{v_t}\alpha _t \right)X_tdt + \alpha _tX_tdW_{1t}-\lambda_t(1+\widetilde{\theta})\mathbb{E}_t\left[I_t(Y)\right]dt\\
        &-\int_{0}^{\infty }\int_{0}^{\infty }R_t(y)\textbf{1}_{\left \{ \lambda\le\lambda_t \right \} }(\lambda)N(dy, d\lambda, dt),\,\,  \\
    \end{aligned}
\end{equation}
where $R_t(Y)=Y-I_t(Y)$ is the retention function. In the last retention term, the inner integral integrates over $y$, and the outer integral integrates over $\lambda$. 

Following existing studies on mean-variance optimization (e.g. \cite{20Zeng2013TimeconsistentIA}, \cite{1Zeng2016robust}, and \cite{3Cao2020OptimalRS}), we assume the individual's objective is to find a sequence of trading and insurance strategies to maximize the following utility function
\begin{equation}\label{problem1}
\mathbb{E}\left [ X_T \right ]-\frac{\gamma}{2} Var\left [ X_T \right ],   
\end{equation}
where $\gamma$ is the risk aversion coefficient. 

\subsection{Orthogonal decomposition and value function}\label{Orthogonal decomposition}
Let $v_{[0, t]}$ and $\lambda_{[0, t]}$ denote paths of $v_t$ and $\lambda_t$ up to time $t$, respectively. An essential difficulty to optimize under current model is that $v_{[0, t]}$ and $\lambda_{[0, t]}$ are not semi-martingales, making the classic Itô formula inapplicable. To circumvent this challenge, we first apply an orthogonal decomposition on $v_{[0, t]}$ and $\lambda_{[0, t]}$, and then use a functional Ito formula to derive the corresponding path-dependent Hamilton–Jacobi–Bellman (P-HJB) equation (see \cite{21article}). Specifically, we decompose the $v_t$ into the following parts
\begin{align}
        &v_s = \Theta_s^t+\zeta_s^t, \,\, 0\le t\le s \le T,\notag\\
        &\Theta_s^t = v_0 +\kappa \int_{0}^{t} K(s-r) (\phi - v_r) dr+\int_{0}^{t} K(s-r) \sigma \sqrt{v_r} d\textbf{B}_r,\label{theta}\\
        &\zeta_s^t= \kappa \int_{t}^{s} K(s-r) (\phi - v_r) dr+\int_{t}^{s} K(s-r) \sigma \sqrt{v_r} d\textbf{B}_r.\notag
    \end{align}
Similarly, we decompose the $\lambda_t$ into the following parts 
\begin{align}
        &\lambda_s = \widetilde{\Theta}_s^t+\widetilde{\zeta}_s^t, \,\, 0\le t\le s \le T,\notag\\
        &\widetilde{\Theta}_s^t = \lambda_0+\int_{0}^{t}\varphi (s-r)(a_0+a_1\lambda_r)dr+\int_{0}^{t}\varphi (s-r)d\pi_r,\label{tilde_theta}\\
        &\widetilde{\zeta}_s^t = \int_{t}^{s}\varphi (s-r)(a_0+a_1\lambda_r)dr+\int_{t}^{s}\varphi (s-r)d\pi_r.\notag
    \end{align}
The $d\pi_t$ above can be written in terms of the Poisson random measure. Therefore, we have an equivalent form
\begin{equation*}
    \begin{aligned}
        &\lambda_s = \widetilde{\Theta}_s^t+\widetilde{\zeta}_s^t, \,\, 0\le t\le s \le T,\\
        &\widetilde{\Theta}_s^t = \lambda_0+\int_{0}^{t}\varphi (s-r)(a_0+a_1\lambda_r)dr+\int_{0}^{t}\int_0^{\infty}\int_0^{\infty}\varphi (s-r)\textbf{1}_{\{\lambda \leq \lambda_r\}}(\lambda)N(dy,d\lambda,dr),\\
        &\widetilde{\zeta}_s^t = \int_{t}^{s}\varphi (s-r)(a_0+a_1\lambda_r)dr+\int_{t}^{s}\int_0^{\infty}\int_0^{\infty}\varphi (s-r)\textbf{1}_{\{\lambda \leq \lambda_r\}}(\lambda)N(dy,d\lambda,dr),\\
    \end{aligned}
\end{equation*}
where in the triple integral, the inner integral integrates over $y$, the middle integral integrates over $\lambda$, and the outer integral integrates over time $r$. Interestingly, $\Theta_s^t$ and $\widetilde{\Theta}_s^t$ are $\mathcal{F}_t$ measurable, and $\zeta_s^t$ and $\widetilde{\zeta}_s^t$ are independent of $\mathcal{F}_t$. This independence is why this method is called ``orthogonal decomposition''. When we take $\mathcal{F}_t$ conditional expectation of the objective, only $\Theta_s^t$ and $\widetilde{\Theta}_s^t$ are left. Consequently, the reward function can be written as a deterministic function of concatenated paths $\omega^1$ and $\omega^2$
\begin{equation*}
J(t, x, \omega^1, \omega^2) = \mathbb{E}_t\left [ X_T \right ]-\frac{\gamma}{2} Var_t\left [ X_T \right ],
\end{equation*}
where $\omega^1$ and $\omega^2$ are obtained by concatenating $\Theta_s^t, \widetilde{\Theta}_s^t$ with
$v_t, \lambda_t$ separately, that is,
$\omega_{s}^1 := v_{s} \mathbf{1}_{\{0 \leq s<t\}}+\Theta_{s}^{t} \mathbf{1}_{\{t \leq s \leq T\}}$, $ \omega_{s}^2 := \lambda_{s} \mathbf{1}_{\{0 \leq s<t\}}+\widetilde{\Theta}_{s}^{t} \mathbf{1}_{\{t \leq s \leq T\}}$. 

\section{Functional Itô formula for the mixed fractional Brownian-Hawkes process}\label{section3:functional formula}
Following  \cite{levental2013simple} and \cite{21article}, we define the sample space in the càdlàg space $D^0$ to include the Hawkes' jumps. Denote
\begin{equation*}
    \begin{aligned}
        &\bar{\Omega}:= D^0([0, T],\mathbb{R}),\,\,\, \Omega_t := D^0([t, T],\mathbb{R}),\,\,\, \bar{\Lambda}:=\left\{(t,\omega)\in [0, T]\times\bar{\Omega}:\omega|_{[t, T]}\in \Omega_t\right\},\\
        &\left \|\omega\right\|_T:=\sup_{0\le t\le T}|\omega_t| ,\,\,\, \textbf{d}((t,\omega),(t^{'},\omega^{'}))=|t-t^{'}|+\|\omega-\omega^{'}\|_T.\\
    \end{aligned}
\end{equation*}
 Let $C^{0}(\bar{\Lambda})$ denote the set of functions $u:\bar{\Lambda}\rightarrow \mathbb{R}$ continuous under $\textbf{d}$. For $u\in C^{0}(\bar{\Lambda})$, the time derivative $\partial_{t}u$ and path derivatives $\partial_{\omega}u, \partial_{\omega\omega}^2u$ are defined as in \cite{21article}, \cite{HanWong}.
 \begin{definition}
     Let $u \in C^0(\bar{\Lambda})$ such that $\partial_\omega u $ exists for all $(t, \omega) \in \bar{\Lambda}$.\\
     (i) We say $\partial_\omega u$ has polynomial growth if there exist constants $C > 0$, $\kappa > 0$ such that
     \begin{equation*}
         \left| \langle \partial_\omega u(t, \omega), \eta \rangle \right| \leq C \left[1 + \|\omega\|_T^\kappa \right] \|\eta \textbf{1}_{[t,T]} \|_{T}, ~\text{for all }~ (t, \omega) \in \bar{\Lambda}, \eta \in \bar{\Omega}.
     \end{equation*}
     (ii) We say $\partial_\omega u$ is continuous if, for any $\eta \in \bar{\Omega}$, the mapping $(t, \omega) \mapsto \langle \partial_\omega u(t, \omega), \eta \rangle$ is continuous under $\textbf{d}$.
 \end{definition}
\begin{definition}\label{defination_3.2}
    A function $u$ is said to belong to $C^{1,2}(\bar{\Lambda})\subset C^{0}(\bar{\Lambda})$ if the derivatives $\partial_t u$, $\partial_{\omega}u$, and $\partial_{\omega\omega}u$ exist and are continuous on $\bar{\Lambda}$. Denote by $C^{1,2}_{+}(\bar{\Lambda})$ the subset of $C^{1,2}(\bar{\Lambda})$ consisting of functions satisfying the following additional conditions:\\
    (i) All the derivatives have polynomial growth.\\
    (ii) The second-order path derivative $\langle \partial_{\omega\omega}^2u ,(\sigma, \sigma) \rangle$ is locally uniformly continuous in $\omega$ and exhibits polynomial growth. Specifically, there exist a constant $\kappa^{'} > 0$ and a bounded modulus of continuity function $\rho$ such that for any $(t, \omega), (t,\omega') \in \bar{\Lambda}$ and $\sigma^{'} \in \Omega_t$,
    \begin{equation*}
        \left|\left \langle  \partial_{\omega\omega}^2u(t,\omega)-\partial_{\omega\omega}^2u(t,\omega^{'}) ,(\sigma^{'}, \sigma^{'})\right \rangle\right|\le\left[1+||\omega||_T^{\kappa^{'}}+||\omega^{'}||_T^{\kappa^{'}}\right]||\sigma^{'}\textbf{1}_{[t, T]}||_T^2\rho\left(||\omega-\omega^{'}||_T\right).
    \end{equation*}
\end{definition}

We consider a mixed fractional Brownian-Hawkes process of the form 
\begin{equation}\label{brownian Hawkes process}
    L_t=l_0+\int_{0}^{t}b(t;r,L_.)dr+\int_{0}^{t}\sigma(t;r,L_.)dW_r+\int_{0}^{t}\int_0^{\infty}\int_{A}\varphi(t;r,L_.)\textbf{1}_{\{\lambda \leq \lambda_r\}}(\lambda)N(dy,d\lambda,dr),
\end{equation}
where $0\le t\le T$, $A$ is bounded below, $W_t$ is a standard Wiener process, $N(dy, d\lambda, dr)$ is the poisson random measure of a compounded Hawkes process, $b$ and $\sigma$ are adapted indicating that $\varphi^{'}(t;r,L_.)=\varphi^{'}(t;r,L_{r\wedge  \cdot})$ for $\varphi^{'}=b, \sigma, \varphi$. 
The corresponding auxiliary processes and paths are
\begin{align*}
    \Theta_s^{t}:= &l_0+\int_{0}^{t}b(s;r,L_.)dr+\int_{0}^{t}\sigma(s;r,L_.)dW_r\\
    &+\int_{0}^{t}\int_0^{\infty}\int_{A}\varphi(s;r,L_.)\textbf{1}_{\{\lambda \leq \lambda_r\}}(\lambda)N(dy,d\lambda,dr) ,
\end{align*}
\begin{equation*}
    \omega_s=(L\otimes_t\Theta^{t})_s:= L_s\textbf{1}_{[0,t)}(s)+\Theta_s^{t}\textbf{1}_{[t,T]}(s),\,\,\,t\le s\le T.
\end{equation*}
We denote the continuous parts of $L_t,\Theta^t_s, \omega_s$ as $(L_t)_c,(\Theta^t_s)_c, (\omega_s)_c$, where
\begin{align}
            &(L_t)_c=l_0+\int_{0}^{t}b(t;r,L_.)dr+\int_{0}^{t}\sigma(t;r,L_.)dW_r
,\,\,\,0\le t\le T,\label{Levy_process}\\
        &(\Theta^{t}_s)_c=l_0+\int_{0}^{t}b(s;r,L_.)dr+\int_{0}^{t}\sigma(s;r,L_.)dW_r,\notag\\
&(\omega_s)_c=(L_t)_c\otimes (\Theta^{t})_c:= (L_s)_c\textbf{1}_{[0,t)}(s)+(\Theta^{t}_s)_c\textbf{1}_{[t,T]}(s),\,\,\,t\le s\le T.\notag            
\end{align}        
In this subsection we assume the following.
\begin{assumption}\label{assumption_1}

        (i) The SDE \eqref{Levy_process} admits a weak solution $((L)_c,W)$.
        
        (ii) $\mathbb{E}\left [ \sup_{0\le t\le T} \left | (L_t)_c \right | ^p \right ]< \infty \,\,\text{for all} \,\,p\ge 1$.
\end{assumption}

\begin{assumption}\label{assumption_2} 
 
$\partial_tb(t;r,\cdot),\partial_{t}\sigma(t;r,\cdot)$ exist for $t\in[r,T]$, and for $\phi = b, \sigma,\partial_tb,\partial_t\sigma$,
    \begin{align*}
        \left | \phi(t;r,\omega) \right | \le C_0\left [ 1+\left \| \omega  \right \|_T^{k_0}  \right ] ,\,\,\, for\,\, some\,\, constants\,\,C_0,k_0>0.
    \end{align*}
\end{assumption}

\begin{assumption}\label{assumption_3} 
    The compensated Hawkes process \\ $\widetilde{N}(dy, d\lambda, dt )=N(dy, d\lambda , dt)-\nu(dy|d\lambda, t) dt$ from \eqref{brownian Hawkes process} is a true martingale. 
\end{assumption}
Under Assumptions \ref{assumption_1} and \ref{assumption_2}, it's obvious that
\begin{align*}
    \mathbb{E}\left [ \left \| (\omega)_c \right \|_T^p \right ]\le C_p, \,\,\, for\,\,all \,\,p\ge1,0\le t\le T.
\end{align*}

Our main result is the following functional Itô formula for the mixed fractional Brownian-Hawkes process.
\begin{theorem}\label{theorem3.1}
    Let Assumptions \ref{assumption_1}, \ref{assumption_2} and \ref{assumption_3} hold and $u\in C_+^{1,2}(\bar{\Lambda})$. Then
    \begin{equation}\label{thm_equation_1}
        \begin{aligned}
            d u(t,\omega) =&\partial _tu(t,\omega)dt+\frac{1}{2}\left \langle \partial^2_{ww}u(t,\omega),(\sigma^{t,L},\sigma^{t,L}) \right \rangle dt\\
            &+\left \langle \partial_{w}u(t,\omega),b^{t,L} \right \rangle dt+\left \langle \partial_{w}u(t,\omega),\sigma^{t,L} \right \rangle dW_t\\
            &+\int_0^{\infty}\int_{A}\left [ u(t,\omega+\varphi^{t,L})-u(t,\omega) \right ]\textbf{1}_{\left\{\lambda\le \lambda_t\right\}}(\lambda)N(dy,d\lambda,dt) ,\\
        \end{aligned}
    \end{equation}
    where $b^{t,L}_s:=b(s;t, L), \sigma^{t,L}_s:=\sigma(s;t, L)$ and $\varphi^{t,L}_s:=\varphi(s;t, L)$ represent the dependence on $s \in [t, T]$.
    \begin{proof} 
    The detailed proof is in Appendix \ref{proof of thm3.1}.
    \end{proof}
\end{theorem}
\begin{remark}
The proof above uses the jump-and-continuous decomposition, a pathwise property that doesn't require Markovianity. One can check it with Section 2.4 and Theorem 3.9 in \cite{friz2018differential}. Moreover, we can also verify the above general result by representing the path with auxiliary state variable $\omega_{s}^1:= v_{s} \mathbf{1}_{\{0 \leq s<t\}}+\Theta_{s}^{t} \mathbf{1}_{\{t \leq s \leq T\}}$, $ \omega_{s}^2:= \lambda_{s} \mathbf{1}_{\{0 \leq s<t\}}+\widetilde{\Theta}_{s}^{t} \mathbf{1}_{\{t \leq s \leq T\}}$ and classical Ito calculus as in Corollary \ref{corollary_4.1}. 
\end{remark}
\section{Optimization problem}\label{sec:solution}

This section derives the optimal strategies under the mean-variance criterion for both the path-dependent and the vanilla cases. It is well known that the dynamic mean‑variance optimization problem \eqref{problem1} is time-inconsistent because the variance term invalidates the classical dynamic programming principle. To address this, we adopt a non-cooperative game perspective, in which one decision maker at each time $t$ is viewed as Player $t$ and competes against each other (see \cite{bjork2017time}, \cite{landriault2018equilibrium}). This leads to the concept of admissible and equilibrium strategies, which we formally define below.
\begin{definition}
    $\textbf{u} =\left \{ (\alpha_t, I_t(\cdot)) \right \}_{t\in [0, T]}$  is said to be an admissible strategy, under the following conditions:
    \begin{enumerate}[label={(\arabic*)}]
        \item $\left\{\textbf{u}_t\right\}_{t\in [0, T]}$ is an adapted process to the filtration $\left\{\mathcal{F}_t\right\}_{t\in [0, T]}$;
        \item $\alpha_t \in \mathbb{R}$ and $I_t(y)\in [0, y]$;
        \item $\mathbb E\left[\int_0^T 
    \int_0^\infty (y-I_t(y))^2\,\mu_{\widetilde Y}(dy)ds+\int_0^T \alpha_s^2ds\right]<\infty$;
        \item Under $\textbf{u}$, the wealth process \eqref{wealth process} admits a unique strong solution.
    \end{enumerate}
    The set of all admissible strategies is denoted by $\mathcal{U}$.
\end{definition}
\begin{definition}
    For an admissible strategy $\textbf{u}^*=\left \{ (\alpha_t^*, I^*_t(\cdot)) \right \}_{t\in [0,T]}$ with any fixed chosen initial state $(t, x, \omega^1_t, \omega^2_t) \in [0, T]\times \mathbb{R} \times \bar{\Omega}\times \bar{\Omega}$ and any admissible strategy $\widetilde{\mathbf u}
=\{(\widetilde{\alpha}_s,\widetilde{I}_s(\cdot))\}_{s\in[t,T]}$, we define the following perturbing strategy
    \begin{equation*}
		\begin{aligned}
			\textbf{u}^{t, h }_s=\begin{cases}
				(\widetilde{\alpha}_s,\widetilde{I}_s(\cdot)), & t\le s< t+h, \\
				\textbf{u}^*_s& t+h \le s \le T.
			\end{cases}
		\end{aligned}
	\end{equation*}
If for every admissible strategy $\widetilde{\mathbf u}$,
    \begin{equation*}
		\lim_{h  \to 0}\inf \frac{J^{ \textbf{u}^*}(t, x, \omega^1, \omega^2)-J^{\textbf{u}^{t,h}}(t, x, \omega^1, \omega^2)}{h} \ge 0.  
    \end{equation*}
    Then $\textbf{u}^*$ is called an equilibrium strategy, and the equilibrium value
    function $F(t, x, \omega^1, \omega^2)$ is defined as $F(t, x, \omega^1, \omega^2)=J^{ \textbf{u}^*}(t, x, \omega^1, \omega^2)$.
\end{definition}

\subsection{Path-dependent case}
In this section, we solve the optimal trading and insurance strategies under the path-dependent case. For the non-Markovian optimization problem, we define
\begin{align*}
	C^{1,2,2,2}&([0,T]\times \mathbb{R}\times \bar{\Omega}\times \bar{\Omega})\\
	&=\left \{ \psi (t,x,\omega^1_t,\omega^2_t)| \psi (t,\cdot,\cdot,\cdot)~ \text{is once continuously differentiable on}~[0,T]\right.\\
	&\left. \text{and} ~\psi (\cdot,x,\omega^1_t, \omega^2_t)~\text{  is twice continuously differentiable on $\mathbb{R}\times \bar{\Omega}\times \bar{\Omega}$ } \right \},\\
	D^{1,2,2,2}&([0,T]\times \mathbb{R}\times\bar{\Omega}\times \bar{\Omega})\\
	&=\left \{ \psi (t,x,\omega^1_t, \omega^2_t)| \psi (t,x,\omega^1_t, \omega^2_t)\in C^{1,2,2,2}([0,T]\times \mathbb{R}\times\bar{\Omega}\times \bar{\Omega})\right.\\
	&\left. \text{and all partial
derivatives satisfy the polynomial growth condition on $\mathbb{R} $} \right \}.
\end{align*}
To obtain the extended HJB equation and the equilibrium strategies, we first derive the infinitesimal generator for the value function.
\begin{theorem}\label{theorem_infitesimal_generator}
    For $\forall (t, x, \omega^1_t, \omega^2_t) \in [0, T]\times \mathbb{R} \times \bar{\Omega}\times \bar{\Omega}, \forall G(t, x, \omega^1_t, \omega^2_t) \in D^{1,2,2,2}([0, T]\times \mathbb{R} \times \bar{\Omega}\times \bar{\Omega})$, $X_t$, $\omega^1$ and $\omega^2$ are defined in \eqref{wealth process} and Section \ref{Orthogonal decomposition}, then the infinitesimal generator is
    \begin{align}
    &\mathcal{A}^{\alpha,I}G(t, X_t, \omega^1_t, \omega^2_t)\notag\\
    =
    &\lim_{h \to 0}\frac{ \mathbb{E}\left[G(t+h,X_{t+h},\omega^1_{t+h}, \omega^2_{t+h})-G(t,X_{t},\omega^1_{t}, \omega^2_{t})|X_t=x\right]}{h}\notag\\
    =&G_t + \Psi_t G_x + \frac{1}{2}\alpha_t^2 x^2 G_{xx} + \left \langle \partial _vG, K(\cdot -t) \Phi _t\right \rangle  +\left \langle \partial _{\lambda}G, \varphi(\cdot -t) (a_0+a_1\lambda_t)\right \rangle\notag\\
    &+\frac{1}{2}\left \langle \partial _{vv}G, (K(\cdot -t)\sigma\sqrt{v_t},K(\cdot -t)\sigma\sqrt{v_t}) \right \rangle +\left \langle \partial _x\partial _vG, K(\cdot -t) \sigma\sqrt{v_t} \right \rangle \rho \alpha_tx \notag\\
    &+\lambda_t\int_0^{\infty}\left[G(t,x-R_t(y),\omega^1_{t},\omega^2_{t} +\varphi(\cdot-t))-G(t,x,\omega^1_{t},\omega^2_{t})\right]\mu_{\widetilde{Y}}(y)dy,\label{infinitesimal_generator}
\end{align}
where 
\begin{align*}
    \Psi_t=\left( \Upsilon + \theta \sqrt{v_t}\alpha _t \right)x-\lambda_t(1+\widetilde{\theta})\mathbb{E}[I_t(Y)],\,\,\,\Phi _t=\kappa (\phi - v_t).
\end{align*}
\begin{proof}
    It is obvious that the volatility $v_t$ in \eqref{volatility_1} and intensity $\lambda_t$ in \eqref{intensity_1} are both mixed fractional Brownian-Hawkes processes with only the fractional Brownian motion part and Hawkes part, respectively. Therefore, we can use the functional Itô formula in Theorem \ref{theorem3.1} to obtain
    \begin{align}
            &dG(t, X_t, \omega^1_{t}, \omega^2_{t})\notag\\
            &=G_tdt + \Psi_t G_x dt+\alpha_txdW_t+ \frac{1}{2}\alpha_t^2 x^2 G_{xx}dt+ \left \langle \partial _vG, K(\cdot -t) \Phi _t\right \rangle dt\notag \\
            &+\left \langle \partial _vG, K(\cdot -t) \sigma\sqrt{v_t}\right \rangle dW_t+\left \langle \partial _{\lambda}G, \varphi(\cdot -t)(a_0+a_1\lambda_t) \right \rangle dt \notag\\
    &+\frac{1}{2}\left \langle \partial _{vv}G, (K(\cdot -t)\sigma\sqrt{v_t},K(\cdot -t)\sigma\sqrt{v_t}) \right \rangle dt+\left \langle \partial _x\partial _vG, K(\cdot -t)\sigma\sqrt{v_t} \right \rangle \rho \alpha_tx dt\notag\\
    &+\int_0^{\infty}\int_{A}\left[ G(t,X_t-R_t(y),\omega^1_{t},\omega^2_{t} +\varphi(\cdot-t))-G(t,X_t,\omega^1_{t},\omega^2_{t})\right]\textbf{1}_{\left\{\lambda\le \lambda_t\right\}}(\lambda)N(dy,d\lambda,dt).\label{infitesimal_generator_expansion}
        \end{align}
According to the definition of the compensated Hawkes process in \eqref{compensated hawkes}, we obtain
\begin{align*}
    &\int_0^{\infty}\int_{A}\left[ G(t,X_t-R_t(y),\omega^1_{t},\omega^2_{t} +\varphi(\cdot-t))-G(t,X_t,\omega^1_{t},\omega^2_{t})\right]\textbf{1}_{\left\{\lambda\le \lambda_t\right\}}(\lambda)N(dy,d\lambda,dt)\\
    &=\int_0^{\infty}\int_{A}\left[ G(t,X_t-R_t(y),\omega^1_{t},\omega^2_{t} +\varphi(\cdot-t))-G(t,X_t,\omega^1_{t},\omega^2_{t})\right]\textbf{1}_{\left\{\lambda\le \lambda_t\right\}}(\lambda)\widetilde{N}(dy,d\lambda,dt)\\
    &+\lambda_t\int_0^{\infty}\left[G(t,x-R_t(y),\omega^1_{t},\omega^2_{t} +\varphi(\cdot-t))-G(t,x,\omega^1_{t},\omega^2_{t})\right]\mu_{\widetilde{Y}}(y)dy.
\end{align*}
After taking the expectation of \eqref{infitesimal_generator_expansion}, we obtain \eqref{infinitesimal_generator}. This completes the proof.
\end{proof}
\end{theorem}

Next, we provide a verification theorem for an equilibrium
insurance-investment strategy for the path-dependent case.
\begin{theorem}\label{verification theorem 4.2}
    (\textbf{Verification Theorem}) For Problem \eqref{problem1}, if there exist two value functions $F(t, x, \omega^1_{t},\omega^2_{t})$ and $g(t, x, \omega^1_{t},\omega^2_{t})\in D^{1,2,2,2}([0, T]\times \mathbb{R} \times \bar{\Omega}\times \bar{\Omega})$ satisfying the following conditions: $\forall (t,x,\omega^1_{t},\omega^2_{t})\in [0,T]\times \mathbb{R} \times \bar{\Omega}\times \bar{\Omega}$,
    \begin{align}
        \sup_{(\mathbf{\alpha}, I)\in \mathcal{U}} &\left \{\mathcal{A}^{\alpha, I}F(t, x, \omega^1_{t},\omega^2_{t})-\mathcal{A}^{\alpha, I}(\frac{\gamma}{2}g^2(t, x, \omega^1_{t},\omega^2_{t}))\right.\notag\\
        &\left.+ \gamma g(t, x, \omega^1_{t},\omega^2_{t})\mathcal{A}^{\alpha, I}g(t, x, \omega^1_{t},\omega^2_{t})\right\}=0,\label{VT11}
        \end{align}
    \begin{equation}\label{VT_F_terminal}
            F(T, x, v_T,\lambda_T) = x,
    \end{equation}
    \begin{equation}\label{VT12}
            \mathcal{A}^{\alpha^{*},I^*}g(t, x, \omega^1_{t},\omega^2_{t}) = 0,\,\,\, g(T, x, v_T,\lambda_T)=x, 
    \end{equation}
    and 
    \begin{align*}
            (\alpha^{*},I^{*}):= \arg\sup_{(\mathbf{\alpha}, I) \in \mathcal{U}} & \left \{\mathcal{A}^{\alpha, I}F(t, x, \omega^1_{t},\omega^2_{t})-\mathcal{A}^{\alpha, I}(\frac{\gamma}{2}g^2(t, x, \omega^1_{t},\omega^2_{t}))\right. \\
            &\left. + \gamma g(t, x, \omega^1_{t},\omega^2_{t})\mathcal{A}^{\alpha, I}g(t, x, \omega^1_{t},\omega^2_{t})\right\}.
        \end{align*}
    Then
    \begin{equation*}
        F(t, x, \omega^1_{t},\omega^2_{t}) = J^{\alpha^{*},I^{*}}(t, x, \omega^1_{t},\omega^2_{t}),\,\,\,
        g(t, x, \omega^1_{t},\omega^2_{t})=\mathbb{E}_t[X^{\alpha^{*}, I^{*}}_T],
    \end{equation*}
    and $\alpha^{*}, I^{*}$ is the equilibrium investment-insurance strategy.
    
    \begin{proof}
        Using the functional Itô calculus rule in Theorem \ref{theorem3.1}, we can easily prove this verification theorem step by step following \cite{HanWong} and \cite{3Cao2020OptimalRS}.
    \end{proof}  
\end{theorem}

Moreover, $\mathbb{E}_t[~\cdot~]$ and $Var_t[~\cdot~]$ are $\mathcal{F}_t$ conditional expectation and variance, respectively.
Due to the special structures of \eqref{volatility_1}  and \eqref{intensity_1}, one can determine the $\mathcal{F}_t$ conditional expectation and variance of the whole path just starting from time $t$, which makes parts  $ v_{s} \mathbf{1}_{\{0 \leq s<t\}}$ and $\lambda_{s} \mathbf{1}_{\{0 \leq s<t\}}$ redundant information in paths $\omega_{s}^1:= v_{s} \mathbf{1}_{\{0 \leq s<t\}}+\Theta_{s}^{t} \mathbf{1}_{\{t \leq s \leq T\}}$ and $ \omega_{s}^2:= \lambda_{s} \mathbf{1}_{\{0 \leq s<t\}}+\widetilde{\Theta}_{s}^{t} \mathbf{1}_{\{t \leq s \leq T\}}$. Therefore, the reward function can be finalized as 
\begin{equation*}
        J(t, x, \omega^1_{t},\omega^2_{t})=J(t, x, \Theta_{[t, T]}^t, \widetilde{\Theta}_{[t, T]}^t)=\mathbb{E}_t\left [ X_T \right ]-\frac{\gamma}{2} Var_t\left [ X_T \right ],
\end{equation*}
where $\Theta_{[t, T]}^t$ and $\widetilde{\Theta}_{[t, T]}^t$ denote the $\Theta_s^t$'s and $\widetilde{\Theta}_s^t$'s paths over $[t, T]$, respectively, see the same form in \cite{21article}, \cite{HanWong}, and \cite{12Han2021RobustCI}. Finally, the individual's value function can be defined as 
\begin{equation*}
F(t, x, \omega^1_{t},\omega^2_{t})=F(t,x,\Theta_{[t,T]}^t,\widetilde{\Theta}_{[t,T]}^t)=\sup_{(\mathbf{\alpha}, I) \in \mathcal{U}} J(t,x,\Theta_{[t,T]}^t,\widetilde{\Theta}_{[t,T]}^t).
\end{equation*}

To solve the HJB equation system in Theorem \ref{verification theorem 4.2}, we assume the functions $F$ and $g$ have the following forms
\begin{align*}
    &F(t, x, \omega^1_{t},\omega^2_{t})=F(t, x, \Theta_{[t,T]}^t,\widetilde{\Theta}_{[t,T]}^t) = A(t) x+ \int_{t}^{T}B(s) \Theta_s^t ds+\int_{t}^{T}C(s)\widetilde{\Theta}_s^t ds+D(t),\\
    &g(t, x, \omega^1_{t},\omega^2_{t})=g(t, x, \Theta_{[t,T]}^t,\widetilde{\Theta}_{[t,T]}^t) = E(t) x+ \int_{t}^{T}H(s) \Theta_s^t ds+\int_{t}^{T}M(s)\widetilde{\Theta}_s^t ds+\breve{N}(t),
\end{align*}
with terminal conditions $A(T)=1$, $D(T)=0$, $E(T)=1$ and $\breve{N}(T)=0$. 

Using Ito calculus and comparing terms of  \eqref{infitesimal_generator_expansion}, the following corollary shows what the path derivatives look like in our case.
\begin{corollary}\label{corollary_4.1}
    Let Assumption \ref{assumption_1}, \ref{assumption_2} and \ref{assumption_3} hold. The functional is
    \begin{equation*}
        f(t,x,\Theta_{[t,T]}^t,\widetilde{\Theta}_{[t,T]}^t)=\textbf{a}(t)x+\int_{t}^{T}\textbf{b}(s) \Theta_s^t ds+\int_{t}^{T}\textbf{c}(s)\widetilde{\Theta}_s^t ds+\textbf{d}(t),
    \end{equation*}
    where $\textbf{a}, \textbf{b},\textbf{c},\textbf{d}$ are bounded deterministic functions, $X_t=x$ is defined as \eqref{wealth process}. $\Theta_{[t,T]}^t$ and $\widetilde{\Theta}_{[t,T]}^t$ are defined as \eqref{theta} and \eqref{tilde_theta}, respectively. Subsequently, the derivatives are given as
    \begin{align*}
        &f_t = \textbf{a}'(t) x - \textbf{b}(t) v_t -\textbf{c}(t)\lambda_t+ \textbf{d}'(t),\,\,\,f_x= \textbf{a}(t), \,\,\, f_{xx}= 0,\\
        &\left \langle \partial _{v}f, K(\cdot -t)\Phi_t \right \rangle = \Phi_t\int_{t}^{T}\textbf{b}(s)K(s-t)ds,\,\,\,\left \langle \partial _{v}f, K(\cdot -t)\sigma\sqrt{v_t} \right \rangle = \sigma\sqrt{v_t}\int_{t}^{T}\textbf{b}(s)K(s-t)ds,\\
    &\left \langle \partial _{x}\partial _{v}f,K(\cdot -t)\sigma\sqrt{v_t}  \right \rangle =\left \langle \partial _{vv}f, (K(\cdot -t)\sigma\sqrt{v_t},K(\cdot -t)\sigma\sqrt{v_t}) \right \rangle =0,\\
    &\left \langle \partial _{\lambda}f, \varphi(\cdot -t)(a_0+a_1\lambda_t) \right \rangle=(a_0+a_1\lambda_t)\int_{t}^{T}\textbf{c}(s)\varphi(s-t)ds,
    \end{align*}
    and the jump increment can be decomposed into two parts: one is reflected in the wealth process, while the other contributes to $\widetilde{\Theta}$, which is
    \begin{align*}
    &\int_0^{\infty}\int_0^{\infty}\left[f(t,x-R_t(y),\Theta_s^t,\widetilde{\Theta}_s^t +\varphi(\cdot-t))-f(t,x,\Theta_s^t,\widetilde{\Theta}_s^t)\right]\textbf{1}_{\{\lambda \leq \lambda_t\}}(\lambda)N(dy,d\lambda,dt)\\
        &=\int_0^{\infty}\int_0^{\infty}\left[-\textbf{a}(t)R_t(y)+\int_{t}^{T}\textbf{c}(s)\varphi (s-t)ds\right]\textbf{1}_{\{\lambda \leq \lambda_t\}}(\lambda)N(dy,d\lambda,dt).
    \end{align*}
    \begin{proof}
        See Appendix \ref{derivatives}
    \end{proof}
\end{corollary}

Denote $\bar{B}_{T-t}:=\int_{t}^{T}B(s)K(s-t)ds$, $\bar{C}_{T-t}:=\int_{t}^{T}C(s)\varphi(s-t)ds$, $\bar{H}_{T-t}:=\int_{t}^{T}H(s)K(s-t)ds$ and $\bar{M}_{T-t}:=\int_{t}^{T}M(s)\varphi(s-t)ds$. The following theorem gives the equilibrium
investment-insurance strategy and the corresponding equilibrium value function.
\begin{theorem}\label{theorem_4.2}
    For the optimal investment-insurance problem \eqref{problem1}
with the wealth process \eqref{wealth process}, the equilibrium strategy is given by
\begin{align*}
    &\alpha_t^*=\frac{\left[\theta -\gamma\sigma\rho\bar{H}_{T-t}\right]\sqrt{v_t}}{A \gamma x}, ~I_t^*(Y)=\left(Y-d^*(t)\right)_+,
\end{align*}
where the optimal deductible $d^*(t)=\left(\frac{\widetilde{\theta}}{\gamma E}+\frac{\bar{M}_{T-t}}{E}\right)_+$, and the corresponding equilibrium value function is
\begin{equation*}
    F(t, x, \Theta_{[t,T]}^t,\widetilde{\Theta}_{[t,T]}^t) = A(t) x+ \int_{t}^{T}B(s) \Theta_s^t ds+\int_{t}^{T}C(s) \widetilde{\Theta}_s^t ds+D(t).
\end{equation*}
In particular,
\begin{align}
        &A=E=e^{\Upsilon(T-t)},\notag\\
        &\bar{B}=K\ast\left[-\kappa\bar{B}-\theta\sigma\rho\bar{H}-\frac{\gamma\sigma ^2(1-\rho^2)}{2}\bar{H}^2+\frac{\theta^2}{2\gamma}\right],\,\,\,\bar{H}=K\ast\left[-(\kappa+\theta\sigma\rho)\bar{H}+\frac{\theta^2}{\gamma}\right],\label{ODE_4.2_B}\\
        &\bar{C}= \varphi\ast \left[g_1(t, \bar{C},\bar{M})\right],\,\,\,\bar{M}= \varphi\ast \left[g_2(t,\bar{M})\right],\label{ODE_4.2_C}\\
        &D=\int_t^T\left(\kappa \phi \bar{B}_{T-s}+a_0\bar{C}_{T-s}\right)ds,\,\,\,\breve{N}=\int_t^T\left(\kappa \phi \bar{H}_{T-s}+a_0\bar{M}_{T-s}\right)ds,\notag
    \end{align}
in which $*$ denotes the convolution operation
\begin{equation*}
    K\ast f(t) =\int_0^tK(t-s)f(s)ds,
\end{equation*}
and  
\begin{align*}
     g_1(t, \bar{C},\bar{M})=&(a_1+1)\bar{C}-(1+\widetilde{\theta})e^{\Upsilon t}\int_{d^*(T-t)}^{\infty}S_{\widetilde{Y}}(y)dy+(\gamma \bar{M}-1)e^{\Upsilon t}\int_0^{d^*(T-t)}S_{\widetilde{Y}}(y)dy\\
        &-\gamma e^{2\Upsilon t}\int_0^{d^*(T-t)}yS_{\widetilde{Y}}(y)dy-\frac{\gamma}{2}\bar{M}^2,\\
        g_2(t,\bar{M})=&(a_1+1)\bar{M}-(1+\widetilde{\theta})e^{\Upsilon t}\int_{d^*(T-t)}^{\infty}S_{\widetilde{Y}}(y)dy-e^{\Upsilon t}\int_0^{d^*(T-t)}S_{\widetilde{Y}}(y)dy,
\end{align*}
where $S_{\widetilde{Y}}$ is the survival function of $\widetilde{Y}$. When $S_{\widetilde{Y}}(y)=e^{-\mu y}$, we have
\begin{align}
        &\bar{C}= \varphi\ast \left[(a_1+1)\bar{C}+\frac{\gamma e^{\Upsilon t}}{\mu}\bar{M}-\frac{\gamma}{2}\bar{M}^2+\frac{\gamma e^{2\Upsilon t}}{\mu^2}e^{-\mu \left(\frac{\widetilde{\theta}}{\gamma e^{\Upsilon t}}+\frac{\bar{M}}{e^{\Upsilon t}}\right)}-\frac{e^{\Upsilon t}}{\mu}-\frac{\gamma e^{2\Upsilon t}}{\mu^2}\right],\\
        &\bar{M}= \varphi\ast \left[(a_1+1)\bar{M}-\frac{\widetilde{\theta}e^{\Upsilon t}}{\mu}e^{-\mu \left(\frac{\widetilde{\theta}}{\gamma e^{\Upsilon t}}+\frac{\bar{M}}{e^{\Upsilon t}}\right)}-\frac{e^{\Upsilon t}}{\mu}\right].\label{M_path-dependent}
\end{align}
\begin{proof}
    See Appendix \ref{proof_thm4.2}.
\end{proof}

\end{theorem}
\subsection{Vanilla case}
For comparison, this section considers the vanilla case in which the path-dependent effects are ignored. Specifically, the intensity of jumps degenerates into the constant $\lambda_0$, and the volatility process degenerates into the classic Heston model
\begin{equation*}\label{volatility_2}
     v_t = v_0 + \kappa\int_{0}^{t} (\phi - v_r) dr + \int_{0}^{t} \sigma \sqrt{v_r} d\textbf{B}_r.
 \end{equation*}
The individual seeks the best choice, i.e., 
\begin{equation}\label{problem2}
    \begin{aligned}
        \sup_{(\mathbf{\alpha}, I)\in\mathcal{U}} J(t,x,v_t; \alpha,I).
    \end{aligned}
\end{equation}
Since the derivation of the vanilla case is trivial, we directly give the following propositions without proof.
\begin{proposition}
    (\textbf{Verification Theorem}) For problem \eqref{problem2}, if there exist two value functions $F(t, x, v_t)$ and $g(t, x, v_t)\in D^{1,2,2}([0, T]\times \mathbb{R} \times \bar{\Omega})$ satisfying the following conditions: $\forall (t,x,v_t)\in [0,T]\times \mathbb{R} \times \bar{\Omega},$
\begin{align*}\label{VT4 1}            \sup_{(\mathbf{\alpha}, I) \in \mathcal{U}}  \left \{\mathcal{A}^{\alpha, I}F(t, x, v_t)-\mathcal{A}^{\alpha, I}\left(\frac{\gamma}{2}g^2(t, x, v_t)\right) + \gamma g(t, x, v_t)\mathcal{A}^{\alpha, I}g(t, x, v_t)\right\}=0,  
\end{align*}
\begin{equation*}
            F(T, x, v_T) = x,
\end{equation*}      
\begin{equation*}\label{VT4 2}
            \mathcal{A}^{\hat{\alpha}, \hat{I}}g(t, x, v_t) = 0, \,\,\,g(T, x, v_T)=x, 
\end{equation*}
    and 
\begin{equation*}
            (\hat{\alpha},\hat{I}):= arg \sup_{(\mathbf{\alpha}, I) \in \mathcal{U}}  \left \{\mathcal{A}^{\alpha, I}F(t, x, v_t)-\mathcal{A}^{\alpha, I}\left(\frac{\gamma}{2}g^2(t, x, v_t)\right) + \gamma g(t, x, v_t)\mathcal{A}^{\alpha, I}g(t, x, v_t)\right\}.
\end{equation*}
    Then
\begin{align*}
        F(t, x, v_t) = J^{\hat{\alpha},\hat{I}}(t, x, v_t),\,\,\,
        g(t, x, v_t)=\mathbb{E}_t[X^{\hat{\alpha},\hat{I}}_T],
\end{align*}
    and $\hat{\alpha},\hat{I}$ is the equilibrium investment-insurance strategy. 
\end{proposition}
The following proposition provides the equilibrium
investment-insurance strategy and the corresponding equilibrium value function.
\begin{proposition}\label{theorem_4.3}
    For the optimal problem \eqref{problem2}
with the wealth process \eqref{wealth process}, the equilibrium strategy is given
by
\begin{align*}
        &\hat{\alpha}_t=\frac{\left[\theta-\sigma\rho\gamma \hat{H} \right]\sqrt{v_t}}{\hat{A}\gamma x}, ~\hat{I}_t(Y)=\left(Y-\hat{d}(t)\right)_+,
    \end{align*}
where the optimal deductible $\hat{d}(t)=\widetilde{\theta}/(\gamma  \hat{E})$, and the corresponding equilibrium value function is
\begin{equation*}
    F(t, x, v_t) = \hat{A}(t) x+ \hat{B}(t)v_t+\hat{D}(t),
\end{equation*}
where 
\begin{align*}
        &\hat{A}=\hat{E}=e^{\Upsilon (T-t)},\,\,\,\frac{\partial \hat{B}}{\partial t}-\kappa \hat{B}-\theta\sigma\rho \hat{H}-\frac{\gamma\sigma ^2(1-\rho^2)}{2}\hat{H}^2+\frac{\theta^2}{2\gamma}=0,\\
        &\frac{\partial \hat{H}}{\partial t}-(\kappa+\theta\sigma\rho)\hat{H}+\frac{\theta^2}{\gamma}=0,\,\,\,\hat{D}=\int_t^Tg_3(s)ds,\,\,\,\hat{N}=\int_t^Tg_4(s)ds,
\end{align*}
in which
\begin{align*}
        g_3(t)=&-(1+\widetilde{\theta})\lambda_0e^{\Upsilon (T-t)}\int_{\frac{\widetilde{\theta}}{\gamma e^{\Upsilon (T-t)}}}^{\infty}S_{\widetilde{Y}}(y)dy+\kappa \phi \hat{B}-\lambda_0e^{\Upsilon (T-t)}\int_0^{\frac{\widetilde{\theta}}{\gamma e^{\Upsilon (T-t)}}}S_{\widetilde{Y}}(y)dy\\
        &-\gamma \lambda_0e^{2\Upsilon (T-t)}
\int_0^{\frac{\widetilde{\theta}}{\gamma e^{\Upsilon (T-t)}}}yS_{\widetilde{Y}}(y)dy,\\
        g_4(t)=&-(1+\widetilde{\theta})\lambda_0e^{\Upsilon (T-t)}\int_{\frac{\widetilde{\theta}}{\gamma e^{\Upsilon (T-t)}}}^{\infty}S_{\widetilde{Y}}(y)dy+\kappa \phi \hat{H}-\lambda_0e^{\Upsilon (T-t)}\int_0^{\frac{\widetilde{\theta}}{\gamma e^{\Upsilon (T-t)}}}S_{\widetilde{Y}}(y)dy,
\end{align*}
where $S_{\widetilde{Y}}(y)$ is the survival function of $\widetilde{Y}$. When $S_{\widetilde{Y}}(y)=e^{-\mu y}$, we have
\begin{align*}
        &\hat{D}=\int_t^T\left(\kappa \phi \hat{B}(s)-\frac{\lambda_0}{\mu}e^{\Upsilon (T-s)}-\frac{\gamma  \lambda_0}{\mu^2}e^{2\Upsilon (T-s)}+\frac{\gamma \lambda_0}{\mu^2}e^{2\Upsilon (T-s)- \frac{\mu\widetilde{\theta}}{\gamma e^{\Upsilon (T-s)}}}\right)ds,\\
        &\hat{N}=\int_t^T\left(\kappa \phi \hat{H}(s)-\frac{\lambda_0}{\mu}e^{\Upsilon (T-s)}-\frac{\widetilde{\theta}\lambda_0}{\mu}e^{\Upsilon (T-s)- \frac{\mu\widetilde{\theta}}{\gamma e^{\Upsilon (T-s)}}}\right)ds.
\end{align*}
\end{proposition}

\section{Numerical illustrations}\label{sec:numerical}
\subsection{Model calibration}
Let $\widetilde{t}_i$ denote the jump times for the Hawkes process $\pi$ such that $0\le \widetilde{t}_1\le \widetilde{t}_2 ~... \le \widetilde{t}_{\widetilde{K}}\le T$. Inspired by \cite{2.hawkes_Patrik}, the likelihood function of the Hawkes process is
\begin{equation*}
    L(\lambda_0,\varrho_1,\varrho_2,p,a_0,a_1)=\left [ \prod_{i=1}^{\widetilde{K}}\lambda_{\widetilde{t}_i}\right ] \exp\left (-\int_{0}^{T}\lambda_s ds \right ). 
\end{equation*}
Subsequently, the log-likelihood function can be written as
\begin{equation}\label{intensity_likelihood}
    l(\lambda_0,\varrho_1,\varrho_2,p,a_0,a_1)=\sum_{j=1}^{\widetilde{K}}\log (\lambda_{\widetilde{t}_j})-\int_{0}^{T} \lambda_s ds,
\end{equation}
where
\begin{equation*}
    \lambda_t =\lambda_0+\int_{0}^{t}(a_0+a_1\lambda_s)\varphi(t-s)ds +\sum_{\widetilde{t}_j<t}\varphi(t-\widetilde{t}_j).
\end{equation*}

We collect the earthquake data from the International Seismological Centre and focus on the 2008 Wenchuan earthquake in Sichuan Province, China ($M_w 7.9$, May 12, 2008, where $M_w$ refers to Moment Magnitude). According to the published report, the Wenchuan earthquake caused 69,227 casualties, 17,923 missing, and 373,643 injuries. The direct economic loss is 845.1 billion dollars (see \cite{ChenBooth2011}). Moreover, a significant aftershock sequence is detected after the Wenchuan earthquake (see \cite{wenchuan}). Inspired by \cite{wenchuan}, we pick up aftershocks with magnitudes greater than 5.0 ($M_w \geq 5.0$ refers to moderate earthquakes causing economic losses) within a 300 km radius of Chengdu city during the period 2008-2023. Table \ref{table1} presents the MLE estimation results of the non-Markovian intensity model and the Markovian intensity model under the subcriticality condition \eqref{subcritical condition} and \eqref{subcritical condition_exponential}  separately. We select the corrected Akaike Information Criterion ($AIC_c$) to evaluate our candidate models because the standard $AIC$ can perform poorly and tend to overfit when the sample size ($n=37$) is small relative to the number of estimated parameters ($k$). As detailed in \cite{burnham2002model}, $AIC_c$ incorporates a crucial second-order bias adjustment and is strongly recommended for proper inference whenever the ratio $n/k$ is less than 40.
\begin{table}[htbp]
\smallskip
\caption{Estimation results of the Hawkes intensity} \label{table1}
\centering % centering table
\begin{tabular}{c r c r}
\multicolumn{4}{c}{Non-Markovian models}\\
\hline
\hline
\multicolumn{4}{c}{Mean-reverting \& power kernel}\\
Log-likelihood &\textbf{8.1208} & $AIC_c$ &-1.4416\\
\hline
Parameter     &Estimate &Parameter     &Estimate \\
\hline
$\lambda_0$ &4.732137& $p$ & 1.001002  \\
$a_0$ & 0.001014 & $a_1$ & -1.644000 \\
$\varrho_1$ & 0.233040&$\varrho_2$ & 0.007060\\
\hline
\hline
\multicolumn{4}{c}{Only power kernel}\\
Log-likelihood &8.0386 & $AIC_c$ &\textbf{-6.8272}\\
\hline
Parameter     &Estimate &Parameter     &Estimate \\
\hline
$\lambda_0$ &1.431227& $p$ & 1.524052  \\
$\varrho_1$ & 0.020010&$\varrho_2$ & 0.003982\\
\hline
\hline
\multicolumn{4}{c}{Markovian models}\\
\hline
\hline
\multicolumn{4}{c}{Mean-reverting \& exponential kernel}\\
Log-likelihood & 6.7185 & $AIC_c$ &-1.5015\\
\hline
Parameter     &Estimate &Parameter     &Estimate \\
\hline
$\lambda_0$ & 0.001020& $a_0$  & 3.214481\\
$a_1$ & -2.339936 & $b_1$  & 4.999991 \\
$b_2$ &0.001184& &\\
\hline
\hline
\multicolumn{4}{c}{Only exponential kernel}\\
Log-likelihood &4.5353 & $AIC_c$ &-2.3433\\
\hline
Parameter     &Estimate &Parameter     &Estimate \\
\hline
$\lambda_0$ &1.185102 & $b_1$ &  2.666847 \\
$b_2$ & 5.000000 &  & \\
\hline
\hline
\end{tabular}
\vspace{2.5ex} % 添加一点表格和注释之间的垂直间距
\begin{minipage}{0.9\textwidth} % 使用 minipage 控制注释的宽度，你可以微调 0.9 这个系数
\footnotesize
The power kernel takes the form $\varphi(t)=\varrho_1/(\varrho_2+t)^p$ and the exponential kernel takes the form $\varphi(t)=b_1e^{-b_2 t}$. ``Only power kernel'' and ``Only exponential kernel'' mean that $a_0=a_1=0$.
\end{minipage}
%\caption*{\footnotesize The power kernel takes the form $\varphi(t)=\varrho_1/\left(\varrho_2+t\right)^p$ and the exponential kernel takes the form  $\varphi(t)=b_1e^{-b_2 t}$. ``Only power kernel'' and ``Only exponential kernel'' mean that $a_0=a_1=0$.}
\end{table}

In Table \ref{table1}, we observe that the ``Mean-reverting \& power kernel'' model has the highest predicting power (highest log-likelihood value). When compared with the ``Only power kernel'' model, we find a trade-off between the mean-reverting pattern and the power decay, as a negative $a_1$ can help the intensity decay. However, mean-reversion introduces two additional parameters and complicates the model. Based on the lowest $AIC_c$ value, we identify the ``only power kernel'' intensity as the optimal choice. Furthermore, ``non-Markovian models'' outperform ``Markovian models'' in both the log-likelihood and the $AIC_c$ values, which show that the power kernel better fits the earthquake data than the exponential kernel. In Table \ref{table1}, the background intensity $\lambda_0 = 1.431227$ represents the base rate of earthquake occurrences unaffected by historical events. The kernel exponent $p=1.524052$ confirms that the decay of aftershock effect follows a power law. The triggering strength $\varrho_1=0.020010$ reflects the immediate increase in future intensity caused by each earthquake, while $\varrho_2=0.003982$ is a small positive constant for model stability. 

\begin{figure}[htbp]\label{intensity_simulation}
        \centering
	{
			\includegraphics[width=4.0in,height=3in]{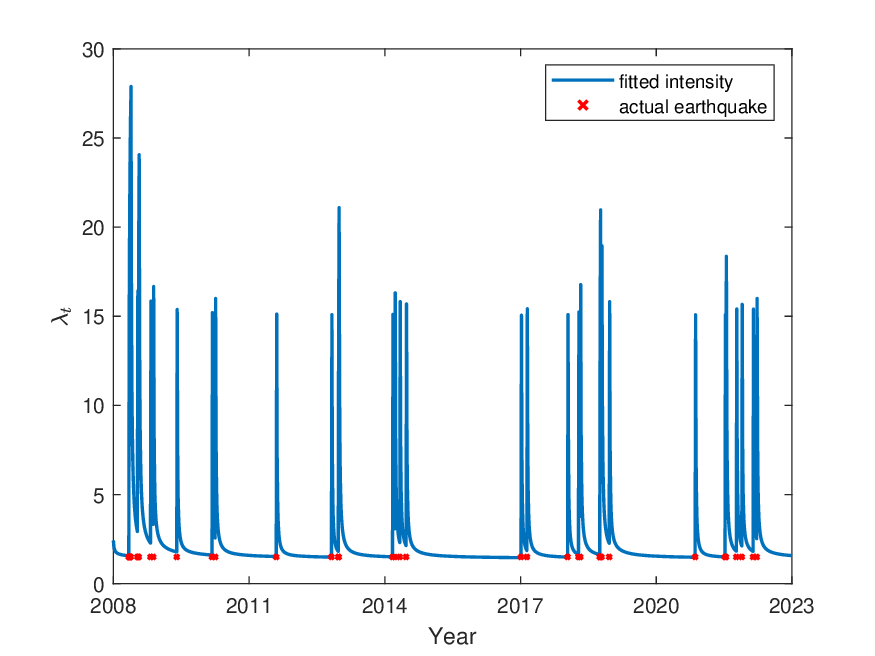}
	}
    \caption{Fitted conditional intensity and observed earthquakes. The solid line is the fitted intensity $\lambda_t$ of the Hawkes process. The dots represent the actual occurrence times of earthquakes with magnitudes $M_w \geq 5.0$ within a 300 km radius of Chengdu city during 2008-2023.}\label{fig1}
\end{figure}

Figure \ref{fig1} displays the estimated conditional intensity (blue solid line) from the Hawkes process model against the observed $M_w \geq 5.0$ magnitude earthquake events (red dots) within a 300 km radius of Chengdu city during 2008-2023. A prominent feature is the intensity peak and the dense clustering of red dots around 2008, which precisely capture the strong self-exciting effect of the Wenchuan earthquake and its subsequent aftershock sequence. Following this, the blue intensity line exhibits an evident power-law decay, consistent with Omori's Law, which describes aftershock activity. Furthermore, several smaller intensity surges and event clusters are visible in later years, reflecting the model's ability to respond to subsequent independent events and their triggered effects.

\subsection{Sensitivity analysis with time horizon}

This section investigates the path-dependent effects on investment and insurance strategies across different time horizons. We assume the losses follow an exponential distribution with survival function $S_{\widetilde{Y}}(y)=e^{-\mu y}$. Our analysis focuses on the two path-dependent parameters: the $\delta \in (\frac{1}{2},1]$ in the kernel function of the rough Heston model $K(t)=t^{\delta-1}/\Gamma(\delta)$ and $p$ satisfies \eqref{subcritical condition} in the kernel function of the Hawkes model $\varphi(t)=\varrho_1/\left(\varrho_2+t\right)^p$. We use the fractional Adams method, as described in \cite{diethelm2004detailed}, to simulate the convoluational ODEs in the explicit solutions. For the financial market, we set the basic parameters from \cite{12Han2021RobustCI}, calibrated on the S\&P 500 implied volatility surface of March 29, 2019 (see Table \ref{table3}). The parameters for the Hawkes process are in Table \ref{table1}'s ``Only power kernel'' case.

\begin{table}[htbp]
\caption{Parameters for financial market under the base scenario} \label{table3}
\smallskip
\centering % centering table
\begin{tabular}{c r c r}
Parameter     &Value &Parameter     &Value\\
\hline
$\kappa$ &0.173 &$\phi$ &0.170\\
$\sigma$ & 0.340 &$\theta$ &5.000\\
$\rho$ &-0.615 &$X_0$ & 1.000\\
$\gamma$ &1.000 &$\mu$ & 4.000\\
$\Upsilon$  &0.020 &$v_0$&0.018\\
$\widetilde{\theta}$ &0.200 &$\delta$ &0.600\\
\hline
\end{tabular}
\end{table}

\begin{figure}[htbp]
    \centering

    \begin{subfigure}[t]{0.48\textwidth}
        \centering
        \includegraphics[width=3.0in,height=2.6in]{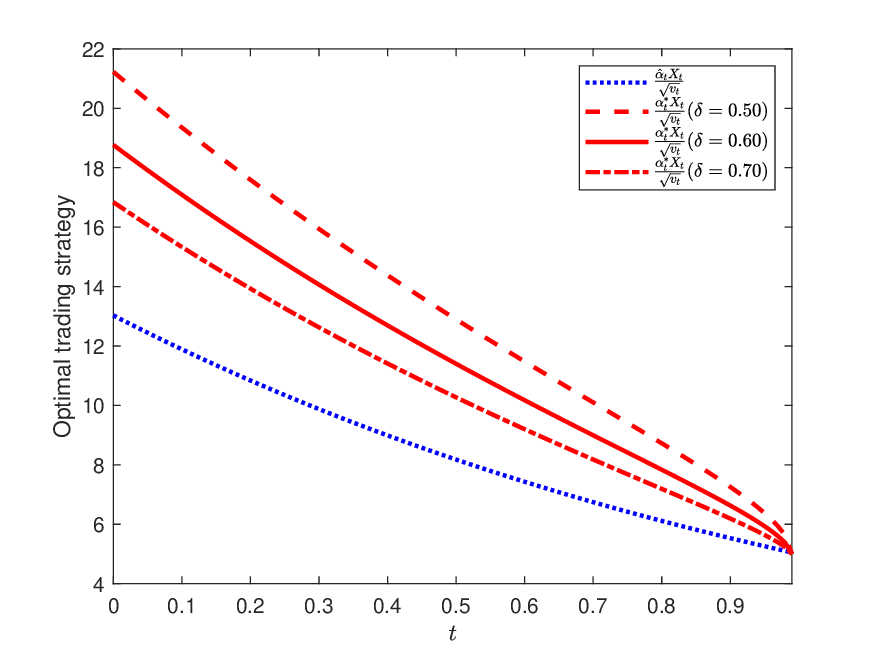}
        \subcaption{$T=1$}
        \label{figure_alpha_a}
    \end{subfigure}
    \hfill
    \begin{subfigure}[t]{0.48\textwidth}
        \centering
        \includegraphics[width=3.0in,height=2.6in]{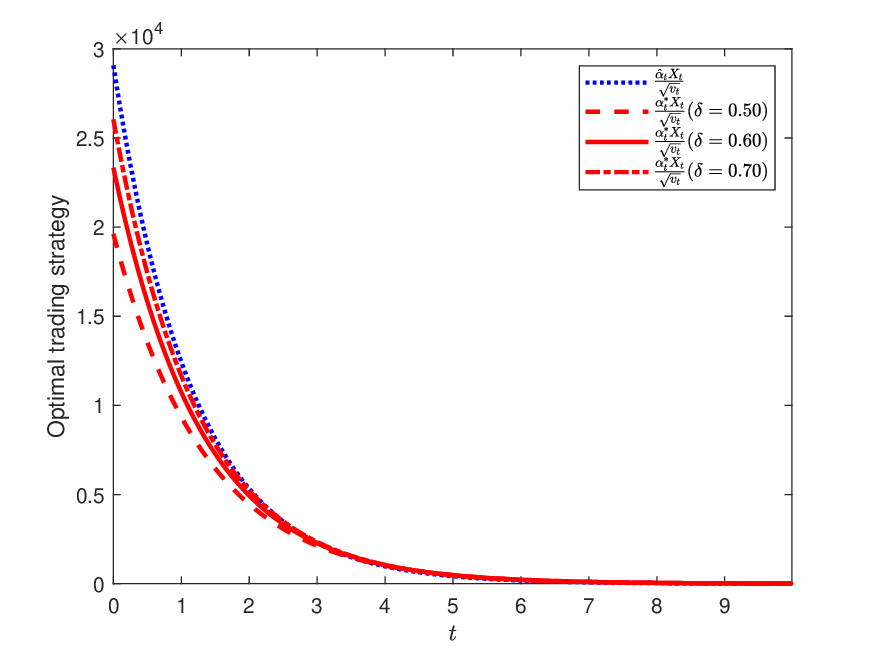}
        \subcaption{$T=10$}
        \label{figure_alpha_c}
    \end{subfigure}

    \caption{Optimal trading strategies with and without path-dependent effects under $\gamma=1$. $T$ is the time horizon. 
    The red lines are the optimal trading strategies $\frac{\alpha_t^* X_t}{\sqrt{v_t}}$ 
    under a different volatility kernel coefficient $\delta$. 
    The blue line is the optimal trading strategy $\frac{\hat{\alpha}_t X_t}{\sqrt{v_t}}$ 
    omitting the path-dependent effect.}
    \label{figures_alpha}
\end{figure}

The optimal investment strategy is highly related to the decision horizon. Figure \ref{figure_alpha_a} demonstrates that over a short-term horizon (e.g., $T = 1$), the path-dependent nature of rough volatility induces risk-seeking behavior compared to the vanilla case. In this scenario, individuals increase their allocation to risky assets to offset potential losses and pursue higher returns. This effect is amplified when volatility is rougher (i.e., $\delta$ is smaller), reflecting the influence of path dependence. Conversely, for a longer-term horizon (e.g., $T=10$), as shown in Figures \ref{figure_alpha_c}, the pattern reverses in the initial phase: rougher volatility leads to lower stock allocations, consistent with the horizon effect documented in Corollary 4.1 of \cite{HanWong}. 

The same phenomenon also occurs in optimal insurance problems. \cite{HanWong} first shows that the effect of roughness reverses with the investment horizon by analyzing the asymptotics of the Mittag-Leffler representation of the strategy. However, the kernel in our model \eqref{M_path-dependent} follows the modified Omori law, which leads to a nonlinear Volterra equation, so no resolvent-type representation is available. Therefore, we prove the following corollary by directly analyzing the derivative of the deductible with respect to $p$.
\begin{corollary}\label{corollary_5.1}
Suppose that $a_1+1\ge 0$ and $p>1$.  Subsequently, the integral $\int_{t}^{T}M(s)\varphi(s-t)ds$ is decreasing with respect to $p$ for $\varrho_2 \to 0$, and increasing for $\varrho_2>1$. 

For $e^{-\frac{1}{p-1}}<\varrho_2<1$, when $T$ is sufficiently large, assume further that 

(i) there exists $\varepsilon\in(0,1)$ such that
\[
|M(t+u)-M(t)|\le \varepsilon |M(t)|,
\,\,\, t\in[0,T],\,\,\, 0\le u\le T-t;
\]

(ii) there exists $Q>0$ such that
\[
\left\|\frac{\partial M}{\partial p}\right\|_\infty\le Q;
\]

(iii) for all $t\in[0,T]$ and $\textbf{J}(t)\ne 0$,
\begin{equation*}
    Q\mathbf A(t)<-M(t)\bigl(|\mathbf J(t)|-\varepsilon \mathbf B(t)\bigr),
\end{equation*}
    where
    \begin{align*}
        &\mathbf A(t):=\int_0^{T-t}(\varrho_2+\tau)^{-p}\,d\tau,
\,\,\,
\mathbf J(t):=\int_0^{T-t}(\varrho_2+\tau)^{-p}\ln(\varrho_2+\tau)\,d\tau,\\
&\mathbf B(t):=\int_0^{T-t}(\varrho_2+\tau)^{-p}\bigl|\ln(\varrho_2+\tau)\bigr|\,d\tau,
    \end{align*}   
      then there exists $t_0\in [0, T]$ such that: $\int_{t}^{T}M(s)\varphi(s-t)ds$ is increasing with respect to $p$ for $t \in [0, t_0)$ and decreasing for $t \in ( t_0, T]$.
\end{corollary}
\begin{proof}
    See Appendix \ref{proof_thm5.1}.
\end{proof}

\begin{figure}[htbp]
    \centering

\begin{subfigure}[t]{0.48\textwidth}
        \centering
        \includegraphics[width=3.0in,height=2.5in]{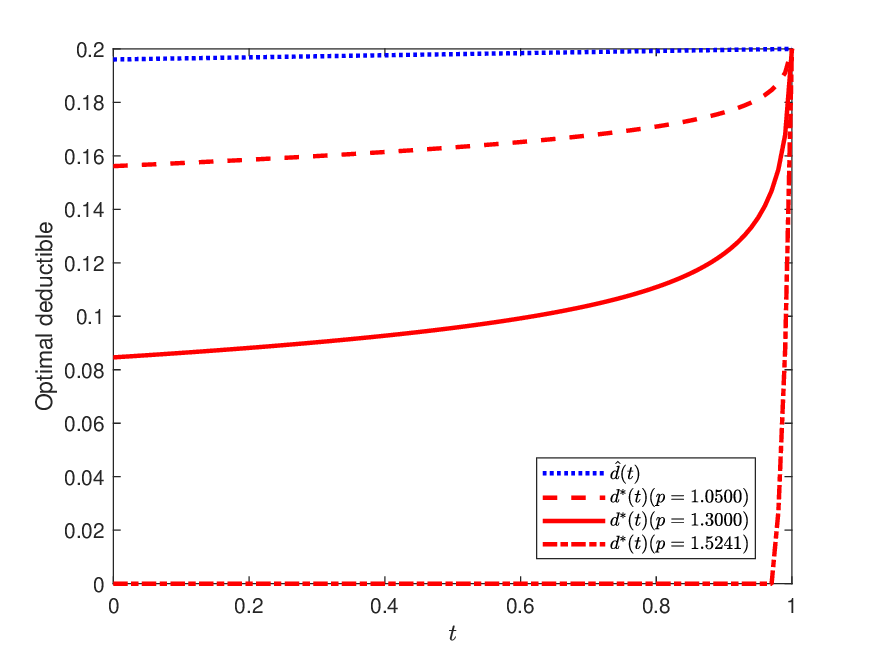}
        \subcaption{$T=1$ , $\varrho_2=0.004$}
        \label{figure_d_a}
    \end{subfigure}
    \hfill
    \begin{subfigure}[t]{0.48\textwidth}
        \centering
        \includegraphics[width=3.0in,height=2.5in]{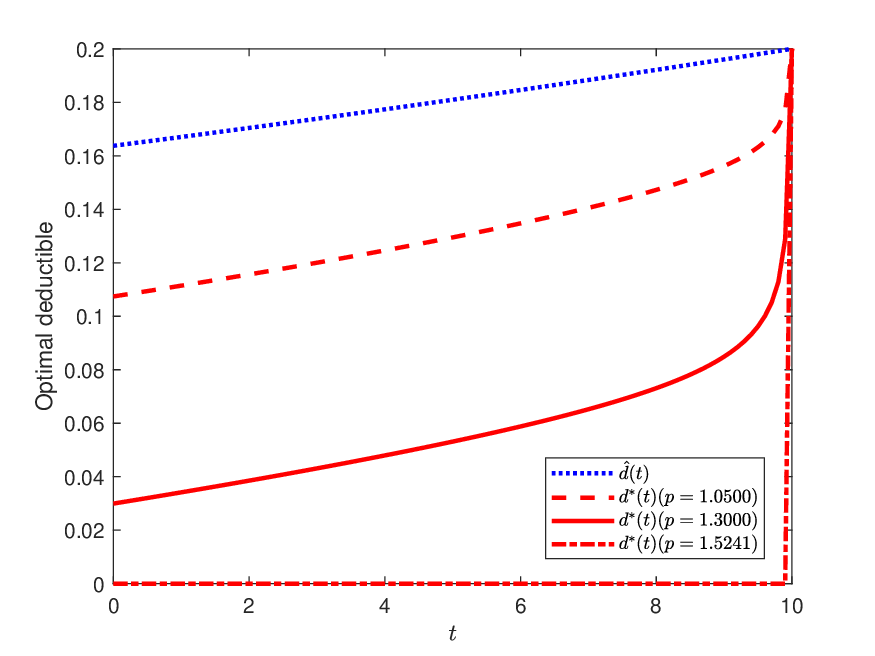}
        \subcaption{$T=10$, $\varrho_2=0.004$ }
        \label{figure_d_b}
    \end{subfigure}
    \hfill
    \begin{subfigure}[t]{0.48\textwidth}
        \centering
        \includegraphics[width=3.0in,height=2.5in]{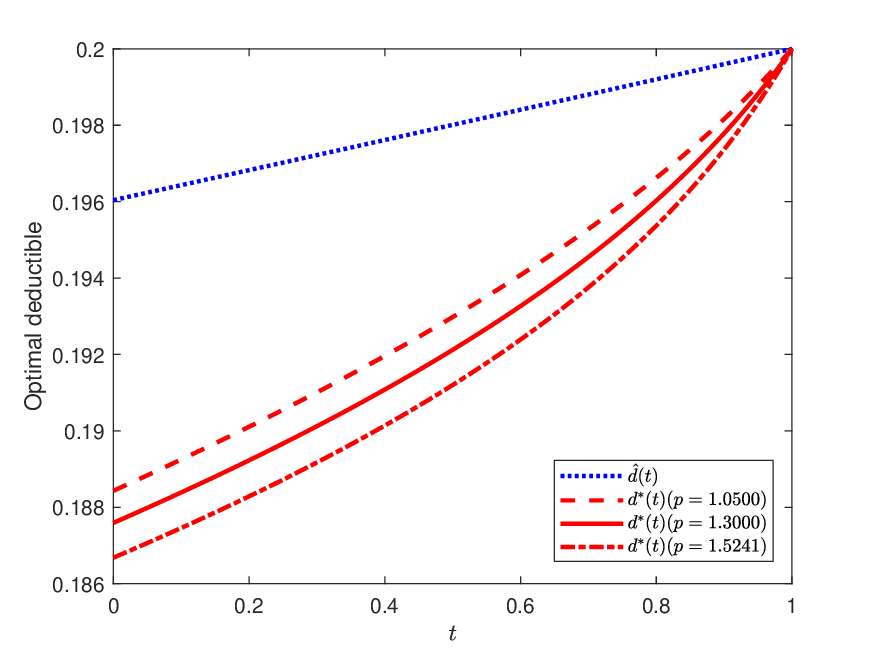}
        \subcaption{$T=1$, $\varrho_2=0.35$}
        \label{figure_d_c}
    \end{subfigure}
    \hfill
    \begin{subfigure}[t]{0.48\textwidth}
        \centering
        \includegraphics[width=3.0in,height=2.5in]{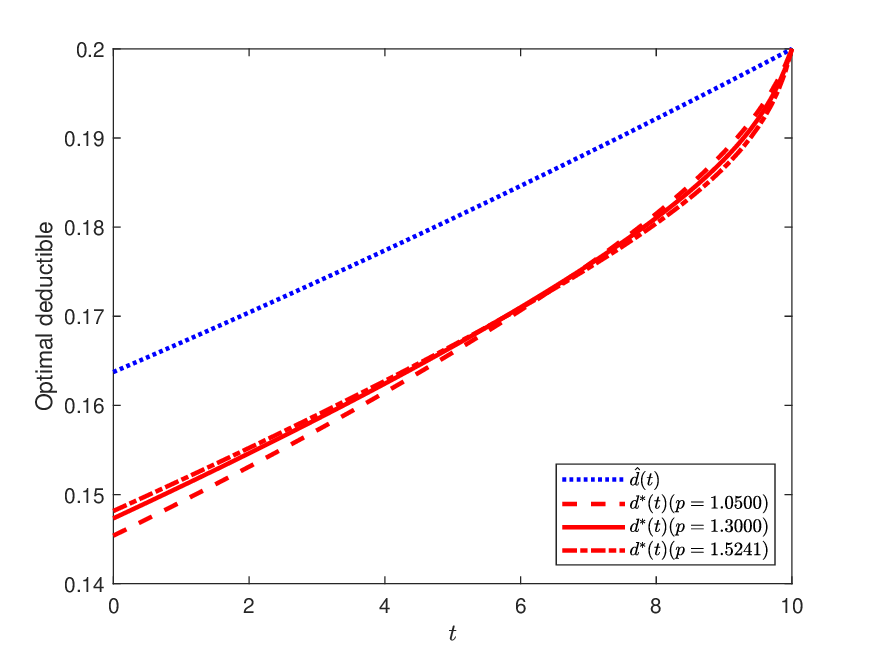}
        \subcaption{$T=10$, $\varrho_2=0.35$}
        \label{figure_d_d}
    \end{subfigure}
    \hfill
    \begin{subfigure}[t]{0.48\textwidth}
        \centering
        \includegraphics[width=3.0in,height=2.5in]{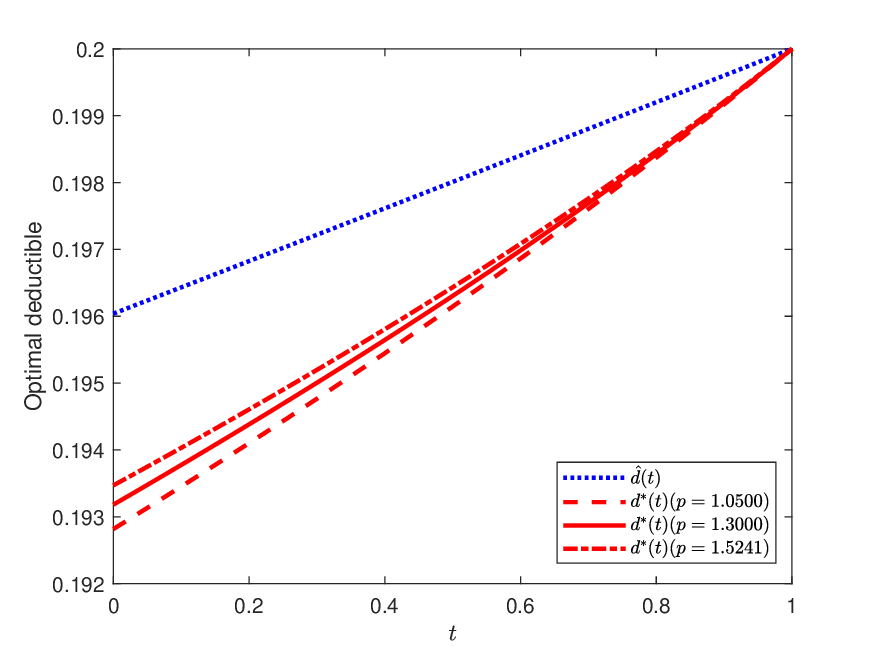}
        \subcaption{$T=1$ , $\varrho_2=1.2$}
        \label{figure_d_e}
    \end{subfigure}
    \hfill
    \begin{subfigure}[t]{0.48\textwidth}
        \centering
        \includegraphics[width=3.0in,height=2.5in]{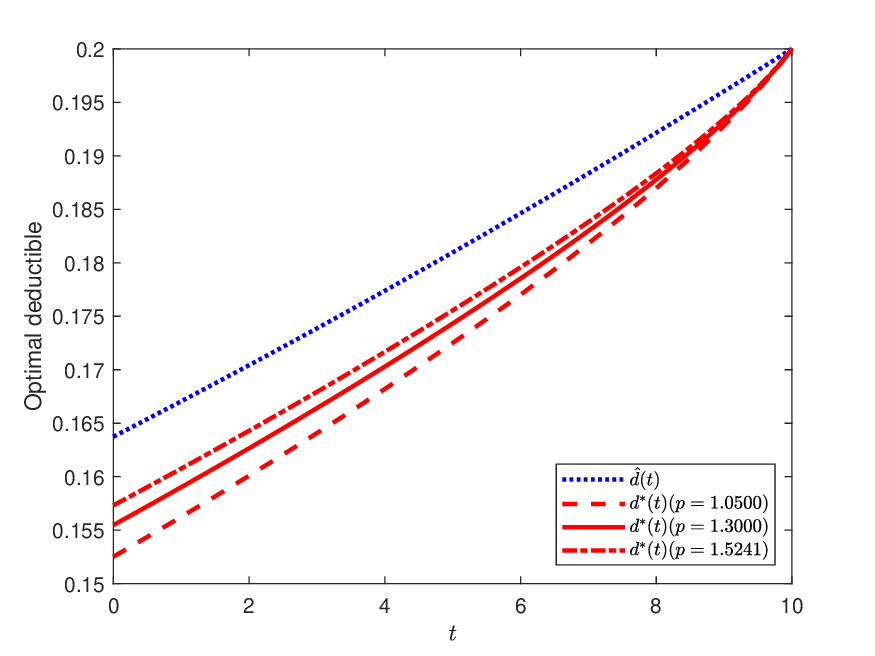}
        \subcaption{$T=10$, $\varrho_2=1.2$ }
        \label{figure_d_f}
    \end{subfigure}

    \caption{Optimal insurance strategies with and without path-dependent effects under $\gamma=1$.}\label{figures_d}
\end{figure}

We next turn to the optimal insurance problem. Figure \ref{figures_d} shows that path-dependent effects of claim arrival intensity amplify the individual's risk-averse behavior relative to the vanilla case, leading to lower deductibles. This is because path dependence propagates previous risks, increasing the current risk level. Figures \ref{figure_d_c} and \ref{figure_d_d} report the case under the parameter $\varrho_2=0.35$. In the short-term horizon (e.g., $T=1$), Figure \ref{figure_d_c} shows that the individual's demand for insurance increases (i.e., lower deductibles) as claim arrival intensity decays more rapidly within the Hawkes process (i.e., $p$ is larger). The reason is that faster decay leads to lower intensity and, consequently, a cheaper premium when severity is fixed. Conversely, in the long-term horizon (e.g., $T=10$), Figure \ref{figure_d_d} reveals an opposite pattern during the early stage: the individual's demand for insurance decreases (i.e., higher deductibles) when the intensity decays faster (i.e., $p$ is larger). Near maturity, the pattern reverts to the short-term horizon case, in which insurance demand increases (i.e., a lower deductible) as intensity decays more quickly. This is consistent with the horizon effect in Corollary \ref{corollary_5.1} for the condition $e^{\frac{-1}{p-1}}<\varrho_2<1$. 

To further investigate the role of $\varrho_2$, Figures \ref{figure_d_a}, \ref{figure_d_b}, \ref{figure_d_e} and \ref{figure_d_f} report the optimal insurance strategies for $\varrho_2=0.004$ and $\varrho_2=1.20$, respectively. Figures \ref{figure_d_a} and \ref{figure_d_b} correspond to the small-$\varrho_2$ condition. In this case, the optimal deductible $d^*(t)$ decreases with respect to $p$ throughout the horizon, indicating that a faster decay of the kernel leads to higher insurance demand. This agrees with the Corollary \ref{corollary_5.1} for the condition $\varrho_2\to 0$. By contrast, Figures \ref{figure_d_e} and  \ref{figure_d_f}  correspond to the $\varrho_2>1$ condition, where a larger $p$ leads to higher deductibles throughout the horizon, indicating a lower demand for insurance. From all things above, Figure \ref{figures_d} provides a numerical illustration of Corollary \ref{corollary_5.1}.

\section{Conclusion}\label{sec:conclusion}

This paper investigates demand for catastrophe insurance under a mean\-variance criterion with path-dependent effects. We consider two non-Markovian models: the rough Heston model for the stock volatility and the Hawkes model for the aftershock intensity. To handle path-dependent effects, we extend the functional Itô calculus formula for fractional Brownian motion to the mixed fractional Brownian-Hawkes process. By introducing auxiliary state variables, we derive path-dependent Hamilton-Jacobi-Bellman (HJB) equations and explicitly solve for the optimal value function and strategies. We calibrate our model using earthquake data from Wenchuan, and the results indicate that the power kernel (Modified Omari Law) fits the data better than the exponential kernel. Numerical simulations reveal that the path-dependent effects are highly related to the time horizon. In investment strategies, rougher stock volatility induces risk-seeking behavior over short-term horizons, whereas under long-term horizons, rougher volatility initially reduces stock allocations, a reversal pattern consistent with the existing literature. In insurance strategy, the path-dependent effects of intensity pass the previous risk to the current level, thereby increasing an individual's demand for catastrophe insurance. In the short term, a faster decay in intensity increases individuals' demand for catastrophe insurance because premiums are cheaper under fixed severity. Conversely, in the long term, an individual's insurance demand decreases as the intensity decays more rapidly, though it reverses to the short-term pattern near maturity. However, we find that this horizon effect vanishes when the shift parameter in the modified Omori kernel approaches zero or exceeds one. Our findings indicate that ignoring path-dependent effects leads to significant underinsurance and highlight the importance of educating customers about this effect in managing catastrophe risk. 

\bibliographystyle{apalike}
\bibliography{ref}

\appendix\label{section:appendix}
    \section{Proof of Proposition \ref{proposition_2.1}}\label{proof of proposition_2.1}
Let $m(t):=\mathbb{E}[\lambda_t],\,\,\,t\ge0$. Since $\pi_t$ has conditional intensity $\lambda_t$, we have
\begin{equation*}
    \mathbb{E}[d\pi_r|\mathcal{F}_{r-}]=\lambda_rdr.
\end{equation*}
Taking expectation on both sides of  \eqref{intensity_1} yields
\begin{equation*}
    m(t)
=
\lambda_0
+\int_0^t \varphi(t-r)a_0\,dr
+\int_0^t \varphi(t-r)(a_1+1)m(r)dr.
\end{equation*}
Integrating over $[0,T]$ and using $\varphi\ge0$, we obtain
\begin{equation*}
    \int_0^T m(t)dt
\le
\lambda_0T+a_0T\|\varphi\|_{L^1(0,T)}
+(a_1+1)_+\|\varphi\|_{L^1(0,T)}\int_0^T m(t)dt.
\end{equation*}
By assumption (ii), we have
\begin{equation*}
    \mathbb{E} \left[\int_0^T \lambda_t dt\right]
=
\int_0^T m(t) dt
\le
\frac{\lambda_0T+a_0T\|\varphi\|_{L^1(0,T)}}
{1-(a_1+1)_+\|\varphi\|_{L^1(0,T)}}
<\infty.
\end{equation*}
Combining this with assumption (i), we have
\begin{equation*}
    \mathbb{E}[\widetilde Y] 
\mathbb{E} \left[\int_0^T\lambda_t dt\right]
<\infty.
\end{equation*}
Thus, the compensated Hawkes process is integrable, hence it is a true martingale on $[0,T]$.

\section{Proof of Theorem \ref{theorem3.1}}\label{proof of thm3.1}
Inspired by the Section 4.4 in \cite{Cambridge_2009}, we first define the jump times for $Y$ as $T_n^A=\inf\left \{ t>T_{n-1}; \bigtriangleup Y(t)\in A\right \}$ for each $n\in \mathbb{N}$ and $T_0=0$. Then we have
        \begin{equation}
            \begin{aligned}
                u(t,\omega)-u(0,0)
                =&\sum_{0\le s\le t} \left [ u(s,\omega_s)-u(s-,\omega_{s-}) \right ]\\ =&\sum_{n=0}^{\infty}\left [ u\left(t\wedge T_{n+1}^A,\omega_{t\wedge T_{n+1}^A}\right)-u\left(t\wedge T_{n}^A,\omega_{t\wedge T_{n}^A}\right) \right ] \\                =&\sum_{n=0}^{\infty}\left [ u\left(t\wedge T_{n+1}^A-,\omega_{t\wedge T_{n+1}^A-}\right)-u\left(t\wedge T_{n}^A,\omega_{t\wedge T_{n}^A}\right)\right ]\\
                &+\sum_{n=0}^{\infty}\left [u\left(t\wedge T_{n+1}^A,\omega_{t\wedge T_{n+1}^A}\right)-u\left(t\wedge T_{n+1}^A-,\omega_{t\wedge T_{n+1}^A-}\right) \right ].\\ 
            \end{aligned}
        \end{equation}

        For the first sum, we know that there are only continuous partial variations of $\omega$ for $\left[t\wedge T_{n}^A,t\wedge T_{n+1}^A-\right]$ by the definition of jump time. Subsequently,
        \begin{equation}
            \begin{aligned}
                &\sum_{n=0}^{\infty}\left [ u\left(t\wedge T_{n+1}^A-,\omega_{t\wedge T_{n+1}^A-}\right)-u\left(t\wedge T_{n}^A,\omega_{t\wedge T_{n}^A}\right)\right ]\\
                &=\sum_{n=0}^{\infty}\left [ u\left(t\wedge T_{n+1}^A,\left(\omega_{t\wedge T_{n+1}^A}\right)_c\right)-u\left(t\wedge T_{n}^A,\left(\omega_{t\wedge T_{n}^A}\right)_c\right)\right ]\\
                &=u(t,\left(\omega_t\right)_c)-u(0,0).\\
            \end{aligned}
        \end{equation}
        Inspired by the Theorem 3.10 in \cite{21article}, we claim that
        \begin{align}
            u\left(t,\left(\omega_t\right)_c\right)-u(0,0)=&\int_0^t\partial _tu(r,(\omega)_c)dr+\frac{1}{2}\int_0^t\left \langle \partial^2_{ww}u(r,(\omega)_c),\left(\sigma^{r,(L)_c},\sigma^{r,(L)_c}\right) \right \rangle dr\notag\\
            &+ \int_0^t\left [ \left \langle \partial_{w}u(r,(\omega)_c),b^{r,(L)_c} \right \rangle dt+\left \langle \partial_{w}u(r,(\omega)_c),\sigma^{r,(L)_c} \right \rangle  \right ]dW_r\notag\\
            =&\int_0^t\partial _tu(r,\omega)dr+\frac{1}{2}\int_0^t\left \langle \partial^2_{ww}u(r,\omega),\left(\sigma^{r,L},\sigma^{r,L}\right) \right \rangle dr\notag\\
            &+ \int_0^t\left [ \left \langle \partial_{w}u(r,\omega),b^{r,L} \right \rangle dt+\left \langle \partial_{w}u(r,\omega),\sigma^{r,L} \right \rangle  \right ]dW_r. \label{functional_Ito_continuous}
        \end{align} 

        For the second sum, there is only a jump that occurs when the interval $\left[t\wedge T_{n+1}-,t\wedge T_{n+1}\right]$ is sufficiently small. Therefore, we have
        \begin{equation*}
            \begin{aligned}
           &\sum_{n=0}^{\infty}\left [u\left(t\wedge T_{n+1},\omega_{t\wedge T_{n+1}}\right)-u\left(t\wedge T_{n+1}-,\omega_{t\wedge T_{n+1}-}\right) \right ]\\
           =&\sum_{n=0}^{\infty}\left [u\left(t\wedge T_{n+1}-,\omega_{t\wedge T_{n+1}-}+\varphi(s;t\wedge T_{n+1},L)\right)-u\left(t\wedge T_{n+1}-,\omega_{t\wedge T_{n+1}-}\right) \right ]\\
           =&\int_{0}^{t}\int_0^{\infty}\int_{A}\left[u\left(r,\omega_r+\varphi(s;r,L)\right)-u\left(r,\omega_r\right)\right] \textbf{1}_{\left\{\lambda\le \lambda_r\right\}}(\lambda)N(dy,d\lambda,dr).\\
            \end{aligned}
        \end{equation*}
        Together with \eqref{functional_Ito_continuous}, we have
        \begin{equation*}
            \begin{aligned}
                u(t,\omega)-u(t,0)=&\int_0^t\partial _tu(r,\omega)dr+\frac{1}{2}\int_0^t\left \langle \partial^2_{ww}u(r,\omega),(\sigma^{r,L},\sigma^{r,L}) \right \rangle dr\\
            &+\int_0^t\left [ \left \langle \partial_{w}u(r,\omega),b^{r,L} \right \rangle dr+\left \langle \partial_{w}u(r,\omega),\sigma^{r,L} \right \rangle  \right ]dW_r\\
            &+\int_{0}^{t}\int_0^{\infty}\int_{A}\left [ u(r,\omega+\varphi^{r,L})-u(r,\omega) \right ]\textbf{1}_{\left\{\lambda\le \lambda_r\right\}}(\lambda)N(dy,d\lambda,dr).\\
            \end{aligned}
        \end{equation*}
    This completes the proof.   
    \section{Proof of Corollary \ref{corollary_4.1}}\label{derivatives}
By the definition of $\Theta_s^t$ and $\widetilde{\Theta}_s^t$ in \eqref{theta} and \eqref{tilde_theta}, we write
\begin{align*}
    Z_t=&\textbf{a}(t)x+\int_{t}^{T}\textbf{b}(s) \Theta_s^t ds+\int_{t}^{T}\textbf{c}(s)\widetilde{\Theta}_s^t ds+\textbf{d}(t)\\
    =&\textbf{a}(t)\left(\int_0^t\Psi_r dr + \int_0^t\alpha _r xdW_{1r}-\int_0^t\int_{0}^{\infty }\int_{0}^{\infty }R_r(y)\textbf{1}_{\left \{ \lambda\le\lambda_r \right \} }(\lambda)N(dy, d\lambda, dr)\right)+\textbf{d}(t)\\
    &+\int_{t}^{T}\textbf{b}(s) \left(v_0 + \int_{0}^{t} K(s-r) \Phi_r dr+\int_{0}^{t} K(s-r) \sigma \sqrt{v_r} d\textbf{B}_r\right) ds\\
    &+\int_{t}^{T}\textbf{c}(s)\left(\lambda_0+\int_{0}^{t}\varphi (s-r)(a_0+a_1\lambda_r)dr\right.\\
    &\left.+\int_{0}^{t}\int_0^{\infty}\int_0^{\infty}\varphi (s-r)\textbf{1}_{\{\lambda \leq \lambda_r\}}(\lambda)N(dy,d\lambda,dr)\right) ds
\end{align*}
We apply the classic It\^{o} formula to $Z_t$, then
\begin{align}
    dZ_t=&\textbf{a}'(t) xdt+\textbf{a}(t)\left(\Psi_t dt + \alpha _t xdW_{1t}-\int_{0}^{\infty }\int_{0}^{\infty }R_t(y)\textbf{1}_{\left \{ \lambda\le\lambda_t \right \} }(\lambda)N(dy, d\lambda, dt)\right)+ \textbf{d}'(t)dt \notag\\
    &-\textbf{b}(t)v_tdt+\int_{t}^{T}\textbf{b}(s) \left(  K(s-t) \Phi_t dt+K(s-t) \sigma \sqrt{v_t} d\textbf{B}_t\right) ds -\textbf{c}(t)\lambda_tdt \notag\\
    &+\int_{t}^{T}\textbf{c}(s)\left( \varphi (s-t)(a_0+a_1\lambda_t)dt+\int_0^{\infty}\int_0^{\infty}\varphi (s-t)\textbf{1}_{\{\lambda \leq \lambda_t\}}(\lambda)N(dy,d\lambda,dt)\right) ds \notag\\
    =&\left(\textbf{a}'(t) x-\textbf{b}(t)v_t-\textbf{c}(t)\lambda_t+\textbf{d}'(t)\right)dt+\textbf{a}(t)\left(\Psi_t dt + \alpha _t xdW_{1t}\right) \notag\\
    &+\left(\int_{t}^{T}\textbf{b}(s) K(s-t)ds\right)\cdot\Phi_t dt+\left(\int_{t}^{T}\textbf{b}(s)K(s-t) ds\right)\cdot\sigma \sqrt{v_t} d\textbf{B}_t \notag\\
    &+\left(\int_{t}^{T}\textbf{c}(s)\varphi (s-t)ds\right)\cdot(a_0+a_1\lambda_t)dt \notag\\
    &+\int_0^{\infty}\int_0^{\infty}\left(-\textbf{a}(t)R_t(y)+\int_{t}^{T}\textbf{c}(s)\varphi (s-t)ds\right)\textbf{1}_{\{\lambda \leq \lambda_t\}}(\lambda)N(dy,d\lambda,dt), \label{Ito_expansion}
\end{align}
where, according to \cite{veraar2012stochastic}, we use the stochastic Fubini theorem to change the order of integrals. Finally, by comparing \eqref{Ito_expansion} and the proof of Theorem \ref{theorem_infitesimal_generator}, we obtain the derivatives and jump increments accordingly. 

\section{Proof of Theorem \ref{theorem_4.2}}\label{proof_thm4.2}
We first calculate the equilibrium strategy $\alpha_t^*$ and $I_t^*$. By the infinitesimal generator \eqref{infinitesimal_generator} and Corollary \ref{corollary_4.1}, we obtain
\begin{align*}
    0=&\frac{d A}{d t} x -B v_t -C \lambda_t +\frac{d D}{d t} +A\Upsilon x+\bar{B}_{T-t}\Phi_t+\bar{C}_{T-t}(a_0+a_1\lambda_t)-\frac{\gamma}{2}\bar{H}_{T-t}^2\sigma ^2v_t\notag\\
    &+\sup_{\mathbf{\alpha} \in \mathbb{R}} \bigg \{\left. A \theta\sqrt{v_t}x\alpha_t -\frac{\gamma}{2}\left[\alpha_t^2x^2E^2+2E\bar{H}_{T-t}\sigma\rho x \alpha_t \sqrt{v_t}\right] \right\}\\
    &+\sup_{I \in [0, y]} \bigg \{\left. -\lambda_tA(1+\widetilde{\theta}) \mathbb{E}[I_t(Y)]\right.\notag\\
    &\left.+\lambda_t\int_0^{\infty}\left[AI_t(y)-\frac{\gamma}{2}\left(2\bar{M}_{T-t}EI_t(y)+E^2I_t(y)^2-2E^2I_t(y)y\right)\right]\mu_{\widetilde{Y}}(y)dy \right\}\\
    &+\lambda_t\int_0^{\infty}\left[-Ay+\bar{C}_{T-t}-\frac{\gamma}{2}\left(2\bar{M}_{T-t}Ey+E^2y^2+\bar{M}_{T-t}^2\right)\right]\mu_{\widetilde{Y}}(y)dy\notag.
\end{align*}
Then we obtain two maximization problems
\begin{align}
    \sup_{\mathbf{\alpha} \in \mathbb{R}} \bigg \{\left. A \theta\sqrt{v_t}x\alpha_t -\frac{\gamma}{2}\left[\alpha_t^2x^2E^2+2E\bar{H}_{T-t}\sigma\rho x \alpha_t \sqrt{v_t}\right] \right\},\label{max_alpha}
\end{align}
and
\begin{align}
    \sup_{I \in [0, y]} &\bigg \{\left. -\lambda_tA(1+\widetilde{\theta}) \mathbb{E}[I_t(Y)]\right.\notag\\
    &\left.+\lambda_t\int_0^{\infty}\left[AI_t(y)-\frac{\gamma}{2}\left(2\bar{M}_{T-t}EI_t(y)+E^2I_t(y)^2-2E^2I_t(y)y\right)\right]\mu_{\widetilde{Y}}(y)dy \right\}\notag\\
    =\sup_{I \in [0, y]} &\bigg \{\left. \lambda_t\int_0^{\infty}\left[-A\widetilde{\theta}I_t(y)-\frac{\gamma}{2}\left(2\bar{M}_{T-t}EI_t(y)+E^2I_t(y)^2-2E^2I_t(y)y\right)\right]\mu_{\widetilde{Y}}(y)dy \right\}.\label{max_I}
\end{align}
According to the first order conditions, differentiating \eqref{max_alpha} with respect to $\alpha$ gives us
\begin{align}
        \alpha_t^*=\frac{\left[\theta -\gamma\sigma\rho\bar{H}_{T-t}\right]\sqrt{v_t}}{A \gamma x}.\label{proof_alphaI}
    \end{align}
 For the problem \eqref{max_I}, the quadratic function
 \begin{align*}
     \widetilde{f}(I_t(y))=-A\widetilde{\theta}I_t(y)-\frac{\gamma}{2}\left(2\bar{M}_{T-t}EI_t(y)+E^2I_t(y)^2-2E^2I_t(y)y\right)
 \end{align*}
 attains the maximum in the interval $I_t(y)\in [0,y]$ at
    \begin{align}
        I_t^*(y)=\left(y-d^*(t)\right)_+,\,\,\, d^*(t)=\left(\frac{\widetilde{\theta}}{\gamma E}+\frac{\bar{M}_{T-t}}{E}\right)_+.\label{proof_I}
    \end{align}
Inserting $\alpha_t^*$ and $I_t^*$ into \eqref{VT11} and \eqref{VT12} yields
\begin{align}
        &\frac{d A}{d t} x - B v_t-C \lambda_t +\frac{d D}{d t}+\bar{C}_{T-t}(a_0+a_1\lambda_t)+\bar{B}_{T-t}\kappa(\phi-v_t)\notag\\
        &+A\left[\Upsilon x+\theta v_t \frac{\left[\theta -\gamma\sigma\rho\bar{H}_{T-t}\right]}{\gamma E}-\lambda_t(1+\widetilde{\theta})\mathbb{E}[I_t^*(Y)] \right]\notag\\
        &-\frac{\gamma}{2}\left[\frac{\left[\theta -\gamma\sigma\rho\bar{H}_{T-t}\right]^2v_t}{\gamma^2}+\bar{H}_{T-t}^2\sigma ^2v_t+2\bar{H}_{T-t}\sigma\rho\frac{\left[\theta-\gamma\sigma\rho\bar{H}_{T-t}\right]v_t}{\gamma}\right]\notag\\
        &+\lambda_t\int_0^{\infty}\left(-R_t^*(y)A+\bar{C}_{T-t}\right)\mu_{\widetilde{Y}}(y)dy-\frac{\gamma}{2}\lambda_t\int_0^{\infty}\left(-R_t^*(y)E+\bar{M}_{T-t}\right)^2\mu_{\widetilde{Y}}(y)dy=0, \label{PDE_1.1} \\
        &\frac{d E}{d t} x - H v_t-M \lambda_t+\frac{d \breve{N}}{d t}+\bar{M}_{T-t}(a_0+a_1\lambda_t)+E\left[\Upsilon x+\theta v_t \frac{\left[\theta  -\gamma\sigma\rho  \bar{H}_{T-t}\right]}{\gamma E } \right.\notag\\
        &\left. -\lambda_t(1+\widetilde{\theta})\mathbb{E}[I_t^*(Y)] \right] +\bar{H}_{T-t}\kappa(\phi-v_t)+\lambda_t\int_0^{\infty}\left(-R_t^*(y)E+\bar{M}_{T-t}\right)\mu_{\widetilde{Y}}(y)dy=0,\label{PDE_1.2}
    \end{align}
where $\Psi_t^*, \Phi_t^*$ and $R_t^*(y)$ denote the quantities obtained by plugging optimal $\alpha_t^*, I_t^*$ into $\Psi_t, \Phi_t$ and $R_t(y)$.
 By separating the variables with and without $x, v_t$ and $\lambda_t$ of \eqref{PDE_1.1} and \eqref{PDE_1.2}, we have
\begin{align}
    A=&E=e^{\Upsilon (T-t)},\label{ODE_1.1_A}\\
        B=&-\kappa\bar{B}_{T-t}-\theta\sigma\rho\bar{H}_{T-t}-\frac{\gamma\sigma ^2(1-\rho^2)}{2}\bar{H}_{T-t}^2+\frac{\theta^2}{2\gamma},\,\,\,
        H=-(\kappa+\theta\sigma\rho)\bar{H}_{T-t}+\frac{\theta^2}{\gamma},\label{ODE_1.1_B}\\
        C=&a_1\bar{C}_{T-t}-A(1+\widetilde{\theta})\mathbb{E}[I_t^*(Y)]+\int_0^{\infty}\left(-R_t^*(y)A+\bar{C}_{T-t}\right)\mu_{\widetilde{Y}}(y)dy\notag\\
        &-\frac{\gamma}{2}\int_0^{\infty}\left(-R_t^*(y)E+\bar{M}_{T-t}\right)^2\mu_{\widetilde{Y}}(y)dy,\label{ODE_1.1_C}\\
        M=&a_1\bar{M}_{T-t}-E(1+\widetilde{\theta})\mathbb{E}[I_t^*(Y)]+\int_0^{\infty}\left(-R_t^*(y)E+\bar{M}_{T-t}\right)\mu_{\widetilde{Y}}(y)dy,\label{ODE_1.1_M}\\
        \frac{d D}{d t}=&-\kappa \phi \bar{B}_{T-t}-a_0\bar{C}_{T-t},\,\,\,
        \frac{d \breve{N}}{d t}=-\kappa \phi \bar{H}_{T-t}-a_0\bar{M}_{T-t}.\label{ODE_1.1_D}
\end{align}
Convolving both sides of $B,H$ in \eqref{ODE_1.1_B} with kernel $K(\cdot)$ and change $T-t$ to $t$ yields
\begin{align}
        &\bar{B}=K\ast\left[-\kappa\bar{B}-\theta\sigma\rho\bar{H}-\frac{\gamma\sigma ^2(1-\rho^2)}{2}\bar{H}^2+\frac{\theta^2}{2\gamma}\right],\label{ODE_BH1.1_B}\\
        &\bar{H}=K\ast\left[-(\kappa+\theta\sigma\rho)\bar{H}+\frac{\theta^2}{\gamma}\right].\label{ODE_BH1.1_H}
\end{align}
By the definition of expectation of $I_t^*=\left(y-d^*(t)\right)_+,\,\,\,d^*(t)=\left(\frac{\widetilde{\theta}}{\gamma E}+\frac{\bar{M}_{T-t}}{E}\right)_+$, we have
\begin{align*}
        &\mathbb{E}[I_t^*(y)]=\int_{d^*(t)}^{\infty}S_{\widetilde{Y}}(y)dy,\,\,\,\int_0^{\infty}R_t^*(y)\mu_{\widetilde{Y}}(y)dy=\int_0^{d^*(t)}S_{\widetilde{Y}}(y)dy,\\
        &\int_0^{\infty}R_t^*(y)^2\mu_{\widetilde{Y}}(y)dy=2\int_0^{d^*(t)}yS_{\widetilde{Y}}(y)dy.
    \end{align*}
Plug them into \eqref{ODE_1.1_C},\eqref{ODE_1.1_M} yields
\begin{align}
        C&=(a_1+1)\bar{C}_{T-t}-A(1+\widetilde{\theta})\int_{d^*(t)}^{\infty}S_{\widetilde{Y}}(y)dy+(\gamma E \bar{M}_{T-t}-A)\int_0^{d^*(t)}S_{\widetilde{Y}}(y)dy\notag\\
        &-\gamma E^2\int_0^{d^*(t)}yS_{\widetilde{Y}}(y)dy
-\frac{\gamma}{2}\bar{M}_{T-t}^2,\label{ODE_CM1.1_C}\\
        M&=(a_1+1)\bar{M}_{T-t}-E(1+\widetilde{\theta})\int_{d^*(t)}^{\infty}S_{\widetilde{Y}}(y)dy-E\int_0^{d^*(t)}S_{\widetilde{Y}}(y)dy.\label{ODE_CM1.1_M}
    \end{align}
Convolving both sides of $C,M$ in \eqref{ODE_CM1.1_C}, \eqref{ODE_CM1.1_M} with kernel $\varphi(\cdot)$ and change $T-t$ to $t$ yields
\begin{align}
        &\bar{C}= \varphi\ast \left[g_1(t, \bar{C},\bar{M})\right],\,\,\,\bar{M}= \varphi\ast \left[g_2(t,\bar{M})\right],\label{ODE_CM1.2_C}
    \end{align}
where
\begin{equation*}
    \begin{aligned}
     g_1(t, \bar{C},\bar{M})&=(a_1+1)\bar{C}-(1+\widetilde{\theta})e^{\Upsilon t}\int_{d^*(T-t)}^{\infty}S_{\widetilde{Y}}(y)dy+(\gamma \bar{M}-1)e^{\Upsilon t}\int_0^{d^*(T-t)}S_{\widetilde{Y}}(y)dy\\
        &-\gamma e^{2\Upsilon t}\int_0^{d^*(T-t)}yS_{\widetilde{Y}}(y)dy-\frac{\gamma}{2}\bar{M}^2,\\
        g_2(t,\bar{M})&=(a_1+1)\bar{M}-(1+\widetilde{\theta})e^{\Upsilon t}\int_{d^*(T-t)}^{\infty}S_{\widetilde{Y}}(y)dy-e^{\Upsilon t}\int_0^{d^*(T-t)}S_{\widetilde{Y}}(y)dy.\\
    \end{aligned}
\end{equation*}
When $F_{\widetilde{Y}}(y)=1-e^{-\mu y}$, then $S_{\widetilde{Y}}(y)=e^{-\mu y}$. \eqref{ODE_CM1.2_C} becomes
\begin{align}
        &\bar{C}= \varphi\ast \left[(a_1+1)\bar{C}+\frac{\gamma e^{\Upsilon t}}{\mu}\bar{M}-\frac{\gamma}{2}\bar{M}^2+\frac{\gamma e^{2\Upsilon t}}{\mu^2}e^{-\mu \left(\frac{\widetilde{\theta}}{\gamma e^{\Upsilon t}}+\frac{\bar{M}}{e^{\Upsilon t}}\right)}-\frac{e^{\Upsilon t}}{\mu}-\frac{\gamma e^{2\Upsilon t}}{\mu^2}\right],\label{ODE_CM1.3_C}\\
        &\bar{M}= \varphi\ast \left[(a_1+1)\bar{M}-\frac{\widetilde{\theta}e^{\Upsilon t}}{\mu}e^{-\mu \left(\frac{\widetilde{\theta}}{\gamma e^{\Upsilon t}}+\frac{\bar{M}}{e^{\Upsilon t}}\right)}-\frac{e^{\Upsilon t}}{\mu}\right].\label{ODE_CM1.3_M}
    \end{align}
Finally, combining \eqref{proof_alphaI}, \eqref{proof_I}, 
\eqref{ODE_1.1_A}, 
\eqref{ODE_1.1_D}-\eqref{ODE_BH1.1_H}, and \eqref{ODE_CM1.2_C}-\eqref{ODE_CM1.3_M}, we prove the theorem.

\section{Proof of Corollary \ref{corollary_5.1}}\label{proof_thm5.1}
Denote $G(t)=\int_{t}^{T}M(s)\varphi(s-t)ds$, then
\begin{equation}
    \begin{aligned}\label{partical of bar_M}
            \frac{\partial G(t)}{\partial p}&= \int_{t}^{T}\left[\frac{\partial M(s)}{\partial p}\varphi(s-t)+M(s)\frac{\partial \varphi(s-t)}{\partial p}\right]ds\\
            &=\int_{t}^{T}\frac{\partial M(s)}{\partial p}\varphi(s-t)ds+\int_{t}^{T}-M(s)\varphi(s-t)\ln(\varrho_2+s-t)ds\\
    \end{aligned}
\end{equation}
Denote $\mathcal{L}(t,\bar{M})=(a_1+1)\bar{M}- \frac{\widetilde{\theta}e^{\Upsilon t}}{\mu}e^{-\mu \left(\frac{\widetilde{\theta}}{\gamma e^{\Upsilon t}}+\frac{\bar{M}}{e^{\Upsilon t}}\right)} - \frac{e^{\Upsilon t}}{\mu}$, from Theorem \ref{theorem_4.2} we have
\begin{equation}\label{M, barM}
    \begin{aligned}
        &M(t)=\mathcal{L}(t, \bar{M}(t)),\,\,\,\bar{M}(t) = \int_{0}^{t} \varphi(t-s) \mathcal{L}(s, \bar{M}(s)) ds.\\
    \end{aligned}
\end{equation}
Since $\bar{M}(0)=0$ and the structure of 
$\mathcal{L}$ ensures non-positivity, we have $M(t)\le 0$. Thus $\bar{M}(t)\le 0$.

If $\varrho_2 \to 0$, denote the first part and second part in \eqref{partical of bar_M} by $\mathcal{I}(t)=\int_{t}^{T}\frac{\partial M(s)}{\partial p}\varphi(s-t)ds$, $\mathcal{D}(t)=\int_{t}^{T}-M(s)\varphi(s-t)\ln(\varrho_2+s-t)ds$. Let $\tau = s-t$, then
\begin{equation*}
    \mathcal{D}(t)=\int_{0}^{T-t}-M(t+\tau)\varphi(\tau)\ln(\varrho_2+\tau)d\tau=\mathcal{D}^+(t)+\mathcal{D}^-(t),
\end{equation*}
where
\begin{equation*}
    \begin{aligned}
        &\mathcal{D}^+(t)=\int_{1-\varrho_2}^{T-t}-M(t+\tau)\varphi(\tau)\ln(\varrho_2+\tau)d\tau\ge0,\\
        &\mathcal{D}^-(t)=\int_0^{1-\varrho_2}-M(t+\tau)\varphi(\tau)\ln(\varrho_2+\tau)d\tau<0.\\
    \end{aligned}
\end{equation*}

Due to the continuity of $M(t)$ on the closed time interval $[0,T]$, there exist constants $M_{\min},M_{\max}>0$ such that $M_{\min}\le |M(t)|\le M_{\max},\,\,\, t\in[0,T]$. Next, we can derive
\begin{equation*}
    \begin{aligned}
        |\mathcal{D}^+(t)|&\le \varrho_1M_{max}\int_{1-\varrho_2}^{T-t}(\varrho_2+\tau)^{-p}\ln(\varrho_2+\tau)d\tau \\
        &\le \varrho_1M_{max}\int_{1}^{\infty}x^{-p}\ln xdx=\frac{\varrho_1M_{max}}{(p-1)^2},\\
        |\mathcal{D}^-(t)|&\ge \varrho_1M_{min}\int_{0}^{1-\varrho_2}(\varrho_2+\tau)^{-p}\ln(\varrho_2+\tau)d\tau \\
        &=\varrho_1M_{min}\left[\frac{1}{(p-1)^2}-\frac{\varrho_2^{1-p}\ln \varrho_2}{p-1}-\frac{\varrho_2^{1-p}}{(p-1)^2}\right].\\
    \end{aligned}
\end{equation*}
When $\varrho_2\to 0$, we have $\varrho_2^{1-p}\to +\infty$ and $\ln \varrho_2\to -\infty$. Thus $|\mathcal{D}^-(t)|\gg |\mathcal{D}^+(t)|$. Consequently, we have $\mathcal{D}(t)<0$.\\
Take partial derivative of $\bar{M}$ in \eqref{M, barM} with respect to $p$ 
\begin{equation*}
    \begin{aligned}
        &\frac{\partial \bar{M}(t)}{\partial p}=\int_{0}^{t} \frac{\partial \varphi(t-s)}{\partial p} \mathcal{L}(s, \bar{M}(s)) ds+\int_{0}^{t}\varphi(t-s) \frac{\partial \mathcal{L}(s, \bar{M}(s))}{\partial \bar{M}}\frac{\partial \bar{M}(s)}{\partial p} ds\\
        &=-\int_{0}^{t} \varphi(t-s)\ln(\varrho_2+t-s)M(s) ds+\int_{0}^{t}\varphi(t-s) \left[a_1+1+\widetilde{\theta}e^{-\mu \left(\frac{\widetilde{\theta}}{\gamma e^{\Upsilon s}}+\frac{\bar{M}}{e^{\Upsilon s}}\right)}\right]\frac{\partial \bar{M}(s)}{\partial p} ds\\
    \end{aligned}
\end{equation*}
Since $\bar{M}(t)$ is predetermined by equation \eqref{M, barM}, then $\frac{\partial \bar{M}(t)}{\partial p}$ satisfies the linear Volterra integral equation of the second kind
\begin{equation}\label{linear volterra equation}
    \begin{aligned}
        \frac{\partial \bar{M}(t)}{\partial p}=\mathcal{G}(t)+\int_{0}^{t}\mathcal{K}(t,s)\frac{\partial \bar{M}(s)}{\partial p} ds\\
    \end{aligned}
\end{equation}
where
\begin{equation*}
    \begin{aligned}
        &\mathcal{G}(t)=-\int_{0}^{t} \varphi(t-s)\ln(\varrho_2+t-s)M(s) ds,\\
        &\mathcal{K}(t,s)=\varphi(t-s) \left[a_1+1+\widetilde{\theta}e^{-\mu \left(\frac{\widetilde{\theta}}{\gamma e^{\Upsilon s}}+\frac{\bar{M}(s)}{e^{\Upsilon s}}\right)}\right].\\
    \end{aligned}
\end{equation*}
Similar to the proof of $\mathcal{D}(t)<0$ above, we can also prove that $\mathcal{G}(t)<0$ when $\varrho_2 \to 0$. Notice that the kernel $\mathcal{K}(t,s)>0$ in \eqref{linear volterra equation}. According to the standard property of Volterra integral equations of the second kind (see \cite{brunner2017volterra}), a positive kernel $\mathcal{K}(t,s)$ generates a positive resolvent kernel $\mathcal{R}(t,s)$. Then the resolvent representation gives 
\begin{equation*}
    \frac{\partial \bar{M}(t)}{\partial p} =\mathcal G(t)
    +
    \int_0^t \mathcal R(t,s)\mathcal G(s)ds.
\end{equation*}
Therefore, the sign of the solution $\frac{\partial \bar{M}(t)}{\partial p}$ follows from the sign of $\mathcal{G}(t)$. Therefore, we conclude that $\frac{\partial \bar{M}(t)}{\partial p}<0$. Thus 
\begin{equation}\label{partical_M_p}
    \begin{aligned}
        \frac{\partial M(t)}{\partial p}=\frac{\partial \mathcal{L}(t, \bar{M}(t))}{\partial \bar{M}}\frac{\partial \bar{M}(t)}{\partial p}=\left[a_1+1+\widetilde{\theta}e^{-\mu \left(\frac{\widetilde{\theta}}{\gamma e^{\Upsilon t}}+\frac{\bar{M}(s)}{e^{\Upsilon t}}\right)}\right]\frac{\partial \bar{M}(t)}{\partial p}<0.
    \end{aligned}
\end{equation}
Thus, the first part in \eqref{partical of bar_M} $\mathcal{I}(t)=\int_{t}^{T}\frac{\partial M(s)}{\partial p}\varphi(s-t)ds<0$. Consequently, we have $\frac{\partial G(t)}{\partial p}<0$. 

If $e^{-\frac{1}{p-1}}<\varrho_2< 1$, denote $\varsigma = T-t$. 
Define
\begin{align*}
    &\textbf{A}(\varsigma):=\int_0^{\varsigma}(\varrho_2+\tau)^{-p}\,d\tau,
\,\,\,
\textbf{J}(\varsigma):=\int_0^{\varsigma}(\varrho_2+\tau)^{-p}\ln(\varrho_2+\tau)\,d\tau,\\
&\textbf{B}(\varsigma):=\int_0^{\varsigma}(\varrho_2+\tau)^{-p}\bigl|\ln(\varrho_2+\tau)\bigr|\,d\tau.
\end{align*}
We write
\begin{equation*}
    \mathcal D(t)
=
-\varrho_1 M(t)\textbf{J}(\varsigma)+R_D(t),
\end{equation*}
where
\begin{equation*}
    R_D(t)
:=
-\varrho_1\int_0^{\varsigma}[M(t+\tau)-M(t)](\varrho_2+\tau)^{-p}\ln(\varrho_2+\tau)\,d\tau.
\end{equation*}
By the assumption (i),
\begin{equation*}
    |R_D(t)|
\le
-\varrho_1\varepsilon M(t)
\int_0^{\varsigma}(\varrho_2+\tau)^{-p}\bigl|\ln(\varrho_2+\tau)\bigr|\,d\tau
=-
\varrho_1\varepsilon M(t)\textbf{B}(\varsigma).
\end{equation*}
By assumption (ii),
\begin{equation*}
    |\mathcal I(t)|
\le \varrho_1 Q \textbf{A}(\varsigma).
\end{equation*}
Hence
\begin{equation*}
    |\mathcal I(t)+R_D(t)|
\le |\mathcal I(t)| +|R_D(t)| \le
\varrho_1 \left[ Q A(\varsigma)-\varepsilon M(t)\textbf{B}(\varsigma)\right].
\end{equation*}
By assumption (iii), we obtain
\begin{equation*}
    |\mathcal I(t)+R_D(t)|
\le 
-\varrho_1 M(t)|\textbf{J}(\varsigma)|.
\end{equation*}
Since 
\begin{equation*}
    \frac{\partial G(t)}{\partial p}=(\mathcal I(t)+R_D(t))-\varrho_1 M(t)\textbf{J}(\varsigma),
\end{equation*}
and $-\varrho_1 M(t)\ge 0$, the term $\mathcal I(t)+R_D(t)$ cannot change the sign of $\frac{\partial G(t)}{\partial p}$. Consequently, the sign of $\frac{\partial G(t)}{\partial p}$ is same as $\textbf{J}(\varsigma)$. Since
\begin{equation*}
    \textbf{J}'(\varsigma)=(\varrho_2+\varsigma)^{-p}\ln(\varrho_2+\varsigma),
\end{equation*}
so that $\textbf{J}'(\varsigma)<0$ for $0<\varsigma<1-\varrho_2$,
$\textbf{J}'(\varsigma)>0$
 for $\varsigma>1-\varrho_2$. Thus $\textbf{J}(\varsigma)$ is strictly decreasing on $(0,1-\varrho_2)$ and strictly increasing on $(1-\varrho_2,\infty)$. Moreover, $\textbf{J}(0)=0$ and $\textbf{J}(\infty)
=
\frac{\varrho_2^{1-p}}{p-1}
\left(\ln\varrho_2+\frac{1}{p-1}\right)>0$ because of $e^{-1/(p-1)}<\varrho_2<1$. Thus, there exists a unique $\varsigma_0>1-\varrho_2$ such that $\textbf{J}(\varsigma_0)=0$. Hence, $\textbf{J}(\varsigma)<0$ for $0\le\varsigma<\varsigma_0$ and 
$\textbf{J}(\varsigma)>0$
 for $\varsigma\ge\varsigma_0$. Consequently, $\frac{\partial G(t)}{\partial p}<0$ for $T-\varsigma_0<t\le T$ and 
$\frac{\partial G(t)}{\partial p}>0$
 for $0\le t<T-\varsigma_0$. Thus, $t_0=T-\varsigma_0$ is the boundary point.

If $\varrho_2> 1$, then $\ln(\varrho_2+t-s)>0$ for all $s, t$. Thus $\mathcal{D}(t) >0$ and $\mathcal{G}(t)> 0$. Similar to the proof when $\varrho_2\to 0$, we have $\frac{\partial \bar{M}(t)}{\partial p} >0$ and $\frac{\partial M(t)}{\partial p} >0$ which lead to $\mathcal{I}(t)>0$. Consequently, we have $\frac{\partial G(t)}{\partial p}>0$.

\end{document}